\documentstyle[12pt]{book}
\newcommand{\bea}{\begin{eqnarray}}
\newcommand{\eea}{\end{eqnarray}}

\newcommand{\pa}{\partial}

\def\bs{b\!\!\!/}

\def\ds{\partial\!\!\!/}
\def\os{\omega\!\!\!\!/}

\begin{document}
\thispagestyle{empty}


\begin{center}
{\LARGE\bf Introduction to modified gravity}

\vspace*{3cm}

{A. Yu. Petrov}

\footnotesize
{\it
 Departamento de F\'{i}sica,\\
Universidade Federal de Paraiba,\\
Jo\~{a}o Pessoa, PB, Brazil}\\

\end{center}

\vfill

This is a preprint of the following work: Albert Petrov, Introduction
  to Modified Gravity, 2020, Springer, reproduced with permission of Springer
  Nature Switzerland AG 2020. The final authenticated version is available
  online at:\\ http://dx.doi.org/10.1007/978-3-030-52862-1 .

\newpage

\thispagestyle{empty}

\tableofcontents

\newpage

\chapter{Einstein gravity and need for its modification}

This review presents itself as a collection of the lecture notes on modified gravity based on lectures given at UFPB (Joao Pessoa, Brazil), CBPF (Rio de Janeiro, Brazil), UFC (Fortaleza, Brazil), and Universidad del Bio-Bio (Concepcion, Chile).

The general relativity (GR) is clearly one of the most successful physical theories. Being formulated as a natural development of the special relativity, it has made a number of fundamental physical predictions which have been confirmed experimentally with a very high degree of precision. Among these predictions, the special role is played by expansion of the Universe and precession of Mercure perihelion, which have been proved many years ago, while other important claims of GR such as gravitational waves and black holes, have been confirmed through direct observations only recently.

By its concept, the general relativity is an essentially geometric theory. Its key idea consists in the fact that the gravitational field manifests itself through modifications of the space-time geometry. Thus, one can develop a theory where the fields characterizing geometry, that is, metric and connection, become dynamical variables so that a non-trivial space can be described in terms of curvature and/or torsion. It has been argued in \cite{PBO} that there are eight types of geometry characterized by possibilities of zero or non-zero curvature tensor, torsion and so-called homothetic curvature tensor, with all these objects are constructed on the base of metric and connection. Nevertheless, the most used formulation of the gravity is based on the Riemannian approach where the connection is symmetric and completely characterized by the metric. Within these lecture notes, we present namely Riemannian description of gravity where the action is described by functions of geometric invariants completely characterized by metric (i.e. various contractions of Riemann curvature tensor, its covariant derivatives and a metric), and possibly some extra fields, scalar or vector ones.  So, let us introduce some basic definitions of quantities used within Riemannian approach.

By definition, the infinitesimal interval in a curved space-time is given as $ds^2=g_{\mu\nu}(x)dx^{\mu}dx^{\nu}$. The metric tensor $g_{\mu\nu}(x)$ is considered as the only independent dynamical variable in our theory. As usual, the action must be (Riemannian) scalar, and for the first step, it is assumed to involve no more than second derivatives of the metric tensor, in a whole analogy with other field theory models where the action involves only up to second derivatives. The only scalar involving only second derivatives is a scalar curvature $R$ (we follow the definitions from the book \cite{Ryder} except of special cases):
\bea
R&=&g^{\mu\nu}R_{\mu\nu}; \quad\, R_{\mu\nu}=R^{\alpha}_{\phantom{\alpha}\mu\alpha\nu};\nonumber\\
R^{\kappa}_{\phantom{\kappa}\lambda\mu\nu}&=&\partial_{\mu}\Gamma^{\kappa}_{\lambda\nu}- \partial_{\mu}\Gamma^{\kappa}_{\lambda\nu}+\Gamma^{\kappa}_{\rho\mu}\Gamma^{\rho}_{\lambda\nu}-\Gamma^{\kappa}_{\rho\nu}\Gamma^{\rho}_{\lambda\mu},
\eea
where $\Gamma^{\mu}_{\nu\lambda}$ are the Christoffel symbols, that is, affine connections expressed in terms of the metric tensor as
\bea
\Gamma^{\mu}_{\nu\lambda}=\frac{1}{2}g^{\mu\rho}(\partial_{\nu}g_{\rho\lambda}+\partial_{\lambda}g_{\rho\nu}-\partial_{\rho}g_{\nu\lambda}).
\eea
The Einstein-Hilbert action is obtained as an integral from the scalar curvature over the $D$-dimensional space-time:
\bea
\label{EH}
S=\int d^Dx\sqrt{|g|}(\frac{1}{2\kappa^2}R+{\cal L}_m),
\eea
where $g$ is the determinant of the metric. We assume the signature to be $(+---)$. The $\kappa^2=8\pi G$ is the gravitational constant (it is important to note that its mass dimension in $D$-dimensional space-time is equal to $2-D$); nevertheless, in some cases we will define it to be equal to 1. The ${\cal L}_m$ is the matter Lagrangian.

Varying the action with respect to the metric tensor, we obtain the Einstein equations:
\bea
\label{ein}
G_{\mu\nu}\equiv R_{\mu\nu}-\frac{1}{2}Rg_{\mu\nu}=\kappa^2 T_{\mu\nu},
\eea
where $T_{\mu\nu}$ is the energy-momentum tensor of the matter. The conservation of the energy-momentum tensor presented as $\nabla_{\mu}T^{\mu\nu}=0$ is clearly consistent with the Bianchi identities $\nabla_{\mu}G^{\mu\nu}=0$.

Among the most important solutions of these equations, one should emphasize the Schwarzschild metric (taking place for the vacuum, $T_{\mu\nu}=0$) which describes the simplest black hole with mass $m$, looking like
\bea
ds^2=(1-\frac{2m}{r})c^2dt^2-(1-\frac{2m}{r})^{-1}dr^2-r^2(d\theta^2+\sin^2\theta d\phi^2)
\eea
(actually, in many cases we will consider a more generic spherically symmetric static metric (\ref{SSSM})), 
and the Friedmann-Robertson-Walker (FRW) metric describing the simplest (homogeneous and isotropic) cosmological solution:
\bea
\label{FRW}
ds^2=c^2dt^2-a^2(t)\left(\frac{dr^2}{1-kr^2}+r^2(d\theta^2+\sin^2\theta d\phi^2)\right),
\eea
where $a(t)$ is the scale factor, and $k=1,0,-1$ for positive, zero and negative curvature respectively. The matter in this case is given by the relativistic fluid:
\bea
\kappa^2T_{\mu\nu}=(\rho+p) v_{\mu}v_{\nu}+p g_{\mu\nu},
\eea
where $\rho$ is a density of the matter, and $p$ is its pressure, in many case one employs the equation of state $p=\omega\rho$, with $\omega$ is a constant characterizing the kind of the matter.

Besides of these solutions, an important example is represented also by the G\"{o}del solution \cite{Godel}:
\bea
\label{Godel}
ds^2=a^2[(dt+e^x dy)^2-dx^2-\frac{1}{2}e^{2x}dy^2-dz^2],
\eea
which, just as the FRW metric, arises if the matter is given by the fluid-like form:
\bea
\kappa^2T_{\mu\nu}=\kappa^2\rho v_{\mu}v_{\nu}+\Lambda g_{\mu\nu},
\eea
but in this case one has $v^{\mu}=\frac{1}{a}$, $\rho=\frac{1}{a^2}$, and $\Lambda=-\frac{1}{2a^2}$. Namely these solutions and their direct generalizations will be considered within our course.

Now, let us make some introduction to quantum gravity. Indeed, it is natural to expect that the gravity, in a whole analogy with electrodynamics and other field theories, must be quantized. To do it, one can follow the approach developed by 't Hooft and Veltman \cite{Veltman}. We start with splitting of the dynamic metric $g_{\mu\nu}$ into a sum of the background part $\bar{g}_{\mu\nu}$ and the quantum fluctuation $h_{\mu\nu}$:
\bea
\label{subs}
g_{\mu\nu}=\bar{g}_{\mu\nu}+\kappa h_{\mu\nu},
\eea
where the $\kappa$ is introduced to change dimension of $h_{\mu\nu}$ to 1.
As a result, the action can be expanded in infinite power series in $h_{\mu\nu}$. For the first step, we can choose $\bar{g}_{\mu\nu}=\eta_{\mu\nu}$.
The lowest, quadratic contribution to the Lagrangian of $h_{\mu\nu}$ is 
\bea
\label{FP}
{\cal L}_0=\frac{1}{4}\partial_{\mu}h_{\alpha}^{\alpha}\partial^{\mu}h_{\beta}^{\beta}-\frac{1}{2}\partial_{\beta}h_{\alpha}^{\alpha}\partial^{\mu}h^{\beta}_{\mu}-\frac{1}{4}\partial_{\mu}h_{\alpha\beta}\partial^{\mu}h^{\alpha\beta}+\frac{1}{2}\partial_{\alpha}h_{\nu\beta}\partial^{\nu}h^{\alpha\beta},
\eea
where the indices of $h_{\alpha\beta}$ are raised and lowered with the flat Minkowski metric. The Lagrangian (\ref{FP}) is called the Fierz-Pauli Lagrangian, it is used within constructing of some generalizations of gravity.

The corresponding (second-order) equations of motion are actually the linearized Einstein equations:
\bea
G^{(0)}_{\mu\nu}\equiv&-&\frac{1}{2}(\pa^{\lambda}\pa_{\mu} h_{\lambda\nu}+\pa^{\lambda}\pa_{\nu}h_{\lambda\mu})+\frac{1}{2}\Box h_{\mu\nu}+\frac{1}{2}\eta_{\mu\nu}\pa_{\alpha}\pa_{\beta}h^{\alpha\beta}-\nonumber\\
&-&\frac{1}{2}\eta_{\mu\nu}\Box h_{\lambda}^{\lambda}+\frac{1}{2}\pa_{\mu}\pa_{\nu} h_{\lambda}^{\lambda}=0.
\eea
We have linearized gauge symmetry $\delta h_{\mu\nu}=\pa_{\mu}\xi_{\nu}+\pa_{\nu}\xi_{\mu}$ in the l.h.s., and linearized Bianchi identities $\pa^{\mu}G^{(0)}_{\mu\nu}=0$. As a consequence, afterwards one must fix the gauge, which can be done by adding the term
\bea
{\cal L}_{GF}=-\frac{1}{2}C_{\mu}C^{\mu},
\eea
where $C_{\mu}=\partial^{\alpha}h_{\alpha\mu}-\frac{1}{2}\partial_{\mu}h_{\alpha}^{\alpha}$, so one has a new Lagrangian
\bea
{\cal L}={\cal L}_0-\frac{1}{2}C_{\mu}C^{\mu}=-\frac{1}{4}\partial_{\mu}h_{\alpha\beta}\partial^{\mu}h^{\alpha\beta}+\frac{1}{8}\partial_{\mu}h_{\alpha}^{\alpha}\partial^{\mu}h_{\beta}^{\beta},
\eea
which can be rewritten as
\bea
{\cal L}=-\frac{1}{2}\partial^{\lambda}h_{\alpha\beta}V^{\alpha\beta\mu\nu}\partial_{\lambda}h_{\mu\nu},
\eea
where $V^{\alpha\beta\mu\nu}=\frac{1}{2}\eta^{\alpha\mu}\eta^{\beta\nu}-\frac{1}{4}\eta^{\alpha\beta}\eta^{\mu\nu}$, which implies the following propagator in the momentum space:
\bea
<h_{\alpha\beta}(-k)h_{\mu\nu}(k)>=i\frac{\eta_{\mu\alpha}\eta_{\nu\beta}+\eta_{\nu\alpha}\eta_{\mu\beta}-\frac{2}{D-2}\eta_{\mu\nu}\eta_{\alpha\beta}}{k^2-i\epsilon},
\eea
where $D$ is the space-time dimension (the singularity at $D=2$ is related with the fact that the $D=2$ Einstein-Hilbert action is a pure surface term).

Now, let us expand the Einstein-Hilbert action (\ref{EH}) in series in $h_{\mu\nu}$ by making again the substitution (\ref{subs}) but with the arbitrary background $\bar{g}_{\mu\nu}$. In this case we see that the metric determinant and curvature scalar are expanded up to the second order in $h$ as
\bea
\label{expand}
\sqrt{|g|}&\to&\sqrt{\bar{|g|}}(1+\frac{1}{2}h_{\alpha}^{\alpha}-\frac{1}{4}h_{\alpha}^{\beta}h^{\beta}_{\alpha}+\frac{1}{8}(h_{\alpha}^{\alpha})^2+\ldots);\\
R&\to& R+\Box h^{\beta}_{\beta}-\nabla^{\alpha}\nabla^{\beta}h_{\alpha\beta}-R^{\alpha\beta}h_{\alpha\beta}-
\frac{1}{2}\nabla_{\alpha}(h_{\mu}^{\beta}h^{\mu,\alpha}_{\beta})+\frac{1}{2}\nabla_{\beta}[h^{\beta}_{\nu}(2h^{\nu\alpha}_{,\alpha}-h_{\alpha}^{\alpha,\nu})]+\nonumber\\&+&
\frac{1}{4}(h^{\nu}_{\beta,\alpha}+h^{\nu}_{\alpha,\beta}-h_{\alpha\beta}^{,\nu})
(h_{\nu}^{\beta,\alpha}+h_{,\nu}^{\beta\alpha}-h_{\nu}^{\alpha,\beta})-\nonumber\\
&-&\frac{1}{4}(2h^{\nu\alpha}_{,\alpha}-h^{\alpha,\nu}_{\alpha})h^{\beta}_{\beta,\nu}-\frac{1}{2}h^{\nu\alpha}h^{\beta}_{\beta,\nu\alpha}+
\frac{1}{2}h^{\nu}_{\alpha}\nabla_{\beta}(h_{\nu}^{\beta,\alpha}+h_{,\nu}^{\beta\alpha}-h_{\nu}^{\alpha,\beta})+h^{\nu}_{\beta}h^{\beta}_{\alpha}R^{\alpha}_{\nu}.\nonumber
\eea
where $h^{\mu,\alpha}_{\beta}\equiv \nabla^{\alpha}h^{\mu}_{\beta}$, etc., and the covariant derivative is constructed on the base of the background metric. This expression is sufficient for the one-loop calculations which yield the following paradigmatic result for the one-loop counterterm arising  from the purely gravitational sector \cite{Veltman}, within the dimensional regularization in $d$-dimensional space-time:
\bea
\label{divs}
\delta {\cal L}=\frac{\sqrt{|g|}}{8\pi^2(d-4)}\left(\frac{1}{120}R^2+\frac{7}{20}R_{\mu\nu}R^{\mu\nu}
\right).
\eea

Many predictions of GR, from expansion of the Universe (which is discussed now in any textbook on general relativity, f.e. in \cite{Ryder}) to existence of gravitational waves whose observations were reported in \cite{LIGO}, have been confirmed through observations.  Nevertheless, it turns out that there are problems which cannot be solved by GR itself, so it requires some modifications. Actually, there are two most important difficulties which the Einstein gravity faced.  The first one is related with the quantum description of the gravity -- indeed, the gravitational constant $\kappa^2$ has a negative mass dimension, precisely to $2-D$ in $D$-dimensional space-time, thus, the Einstein-Hilbert gravity is non-renormalizable, i.e. its consistent description must involve an infinite number of counterterms (an excellent review on quantum calculations in gravity is presented in the book \cite{BOS}). The second difficulty consists in the fact that  the cosmic acceleration whose discovery  was reported in \cite{Riess} has not been predicted theoretically since it does not admit explanations within the general relativity. 

Therefore the problem of possible modifications of gravity arises naturally. Actually, although first attempts to introduce modified gravity have been carried out much earlier, these two discoveries increased radically attention to modified gravity models.

The simplest attempt to solve the cosmic acceleration problem is based on the introducing the cosmological constant $\Lambda$, i.e. we add to the action (\ref{EH}) the extra term $S_{\Lambda}=-\frac{1}{\kappa^2}\Lambda\int d^4x \sqrt{|g|}$, so, in the l.h.s. of (\ref{ein}), the additive term $\Lambda g_{\mu\nu}$ will arise. 
It is easy to see that for the FRW metric, the components of the Ricci tensor and the scalar curvature are
\bea
\label{cosmcurv}
R_{00}&=&-\frac{3\ddot{a}}{a};\quad\, R_{ij}=\delta_{ij}(a\ddot{a}+2\dot{a}^2);\nonumber\\
R&=&6\left(\frac{\ddot{a}}{a}+\frac{\dot{a}^2}{a^2}+\frac{k}{a^2}
\right).
\eea
For the FRW metric (\ref{FRW}), the Einstein equation for the (00) component, together with the equation obtained as difference of (ii) and (00) equations, with $c=1$ and $\kappa^2=8\pi G$, yield
\bea
&&\frac{\dot{a}^2}{a^2}+k=\frac{1}{3}(8\pi G\rho+\Lambda);\\
&&\frac{\ddot{a}}{a}=-\frac{4}{3}\pi G(\rho+3p)+\frac{\Lambda}{3},\nonumber
\eea
where $k=+1,0,-1$ for positive, zero and negative scalar curvature. As it is well known, originally $\Lambda$ was introduced by Einstein in order to provide a static solution while further de Sitter proved that the empty space with negative $\Lambda$ will expand exponentially. Therefore, after discovery of the cosmic acceleration  the idea of the cosmological constant has been revitalized \cite{Carroll}. However, the cosmological constant, by astronomical observations, should be extremely small (about 120 order less than a natural scale for it given by $M_{Planck}^4$), and this fact has no theoretical explanation (the search for this explanation constitutes the famous cosmological constant problem). Besides, the cosmological constant does not solve the problem of renormalizability of gravity. 

There are two manners how to extend the gravity in order to solve these problems. Within the first approach, we modify the Einstein-Hilbert action through introducing additive terms. Within the second approach, we suggest that the full description of gravity involves, besides of the metric field, also some extra scalar or vector fields which must not be confused with matter being treated as ingredients of the gravity itself, so that usual results of Einstein gravity are recovered, for example, when these fields are constant (the typical example is the Brans-Dicke gravity which we discuss further). In this review, we give a description of these approaches. It should be noted that among these approaches, an important role is played by adding new terms (and/or fields) aimed either to break the Lorentz/CPT symmetry or to introduce a supersymmetric extension of gravity. Within this review we also discuss these approaches.

The structure of this review looks like follows. In the chapter 2, we present various models obtained through modifications of the purely gravitational sector. In the chapter 3, we consider various scalar-tensor gravity models, such as Chern-Simons and Brans-Dicke gravities, and galileons. In the chapter 4, we discuss vector-tensor gravity models and problem of Lorentz symmetry breaking in gravity. In the chapter 5, we review most interesting results in Horava-Lifshitz gravity. In the chapter 6, we discuss some results for nonlocal gravity. The chapter 7 represents conclusions of our course.

\chapter{Modifications of the pure gravitational sector}

\section{Motivations}

As we already noted in the Introduction, one of the ways to modify gravity consists in introducing additional terms to the gravitational sector. Such terms are given by scalars constructed on the base of the metric tensor, i.e. these scalars are functions of the Riemann tensor, the Ricci tensor, possibly, their covariant derivatives, and the scalar curvature. In the simplest case the Lagrangian is the function of the scalar curvature only, so, the action is
\bea
\label{fr}
S=\frac{1}{16\pi G}\int d^4x \sqrt{|g|}f(R),
\eea
where, $f(R)$ is a some function of the scalar curvature. Since the Einstein gravity is very well observationally confirmed, and the curvature of the Universe is known to be small, it is natural to suggest that $f(R)=R+\gamma R^n$, with $n\geq 2$, so, Einstein-Hilbert term dominates.  The case $n=2$ is very interesting by various reasons, from renormalizability to possibility of cosmic acceleration, so, it will be discussed in details. However, other values of $n$, including even negative ones which called attention recently, are also interesting. Another generalization of this action is the suggestion that the Lagrangian depends also on invariants $Q=R_{\mu\nu}R^{\mu\nu}$ and $P=R_{\mu\nu\alpha\beta}R^{\mu\nu\alpha\beta}$, such class of theories is called $f(R,Q,P)$ gravity, the paradigmatic example is the Weyl gravity (see f.e. \cite{Ghile} and references therein), where the Lagrangian is given by the square of the Weyl tensor. 
Besides of these situations, it is interesting also to abandon the restriction for the space-time to be four-dimensional. In this context we will consider also higher-dimensional space-times and discuss Lovelock gravities whose action involves higher curvature invariants.

\section{$R^2$-gravity}

Let us start with the action
\begin{equation}
\label{r2}
S=\frac{1}{16\pi G}\int d^4x\sqrt{|g|}(R+\alpha R_{\mu\nu}R^{\mu\nu}-\beta R^2)+S_{mat}.
\end{equation}
A simple comparison of this expression with (\ref{expand}) shows that this action is of the second order in curvatures, i.e. of fourth order in derivatives, therefore the theory described by this action is called $R^2$-gravity.  In principle, one can add also the square of the Riemann tensor, however, since in  the four-dimensional space-time the Gauss-Bonnet term ${\cal G}=R^2-4R_{\mu\nu}R^{\mu\nu}+R_{\mu\nu\lambda\rho}R^{\mu\nu\lambda\rho}$ is a total derivative, the square of the Riemann tensor in $D=4$ is not independent.

We see that the additive term in this action exactly matches the structure of the one-loop divergence arising in the pure Einstein gravity (\ref{divs}). Therefore, the theory (\ref{r2}) is one-loop renormalizable. Moreover, it is not difficult to show that no other divergences arise in the theory. Here, we demonstrate it in the manner similar to that one used within the background field method for the super-Yang-Mills theory \cite{West}. Indeed, the propagator in this theory behaves as $k^{-4}$. Any vertex involves no more than four derivatives. Integration over internal momentum in any loop yields the factor 4, hence formally the superficial degree of divergence must be $\omega=4L-4P+4V=4$. However, we should take into account that this is the upper limit for $\omega$, and each derivative acting to the external legs instead of the propagator decreases $\omega$ by 1. Since $R_{\mu\nu\lambda\rho}$, as well as the Ricci tensor, involves second derivatives, each external $R_{\mu\nu\lambda\rho}$, $R_{\mu\nu}$, $R$ decreases the $\omega$ by 2. Hence, the $R^2$ or $R_{\mu\nu}R^{\mu\nu}$ contributions will display only logarithmic divergences, and higher-order contributions like $R^3$ will yield $\omega<0$ being thus superficially finite. The presence of Faddeev-Popov (FP) ghosts does not jeopardize this conclusion since their Lagrangian looks like \cite{Veltman}
\bea
{\cal L}_{gh}=\bar{C}_{\rho}\delta^{\rho}_{\mu}\pa_{\nu}(D_{\alpha}^{\mu\nu}C^{\alpha}),
\eea
where $C$, $\bar{C}$ are the FP ghosts, and $D^{\mu\nu}$ is the operator defined from gauge transformations for the metric fluctuation $h^{\mu\nu}$:
$$
D^{\mu\nu}_{\alpha}\xi^{\alpha}\equiv\pa^{\mu}\xi^{\nu}+\pa^{\nu}\xi^{\mu}-\eta^{\mu\nu}\pa_{\alpha}\xi^{\alpha}+\pa_{\mu}\xi_{\alpha} h^{\alpha\nu}+ \pa_{\nu}\xi_{\alpha} h^{\alpha\mu}+
\xi^{\alpha}\pa_{\alpha} h^{\mu\nu}-\pa_{\alpha}\xi^{\alpha} h^{\mu\nu}.
$$
So, the propagator of ghosts is proportional to $k^{-2}$, while the vertex contains only one derivative. Clearly, presence of ghosts will decrease the $\omega$.

Let us discuss various aspects of the theory (\ref{r2}). We follow the argumentation presented in \cite{Stelle,Stelle1}. First, one can write down the equations of motion:
\bea
\label{hmn}
H_{\mu\nu}&\equiv& (\alpha-2\beta)\nabla_{\mu}\nabla_{\nu}R-\alpha \Box R_{\mu\nu}-(\frac{\alpha}{2}-2\beta)g_{\mu\nu}\Box R+\nonumber\\ &+&
2\alpha R^{\rho\lambda}R_{\mu\rho\nu\lambda}-
2\beta RR_{\mu\nu}-\frac{1}{2}g_{\mu\nu}(\alpha R^{\rho\lambda}R_{\rho\lambda}-\beta R^2)+
\nonumber\\ &+& \frac{1}{G}(R_{\mu\nu}-\frac{1}{2}Rg_{\mu\nu})=T_{\mu\nu}.
\eea
Using these equations, one can find the Newtonian static limit of the theory. Proceeding in the same way as in GR, we can show that the gravitational potential in the non-relativistic limit is
\bea
\label{pot}
\phi=h^{00}=\frac{1}{r}-\frac{4}{3}\frac{e^{-m_2r}}{r}+\frac{1}{3}\frac{e^{-m_0r}}{r},
\eea
where $m_0=(16\pi G\alpha)^{-1/2}$, and $m_2=(32\pi G(3\beta-\alpha))^{-1/2}$. So, we find that the $R^2$-gravity involves massive modes displaying Yukawa-like contributions to the potential. Following the estimations from \cite{Stelle}, the $m_{0,2}$ are about $10^{-17}$ $M_{Pl}$. We note that the Birkhoff theorem is no more valid in this theory since there are mass-like parameters $m_0$, $m_2$, and instead of the Bianchi identities one will have $\nabla_{\mu}H^{\mu\nu}=0$.

Then, it is interesting to discuss cosmological solutions in this theory. A remarkable feature of the $R^2$-gravity consists in the fact that it was the first gravity model to predict accelerated expansion of the Universe much before its observational discovery. The pioneer role was played by the paper \cite{Staro}, where terms of higher orders in curvature generated by some anomaly have been introduced to the equation of motion, so the resulting equation, for the vacuum, looks like
\bea
\label{sta}
G_{\mu\nu}&=&k_1(R_{\mu}^{\lambda}R_{\nu\lambda}-\frac{2}{3}RR_{\mu\nu}-\frac{1}{2}g_{\mu\nu}R_{\alpha\beta}R^{\alpha\beta}+\frac{1}{4}g_{\mu\nu}R^2)+\nonumber\\
&+&k_2(\nabla_{\nu}\nabla_{\mu}R-2g_{\mu\nu}\Box R-2RR_{\mu\nu}+\frac{1}{2}g_{\mu\nu}R^2),
\eea
where $k_1,k_2$ are constants. Many terms in the r.h.s. of this equation are present also in (\ref{hmn}), actually, at $\alpha=0$ and $k_1=0$ these equations coincide up to some numerical coefficients, so, their solutions are not very different. Substituting the FRW metric into (\ref{sta}), we arrive at
\bea
\label{sta1}
\frac{\dot{a}^2+k}{a^2}&=&\frac{1}{H^2}\left(\frac{\dot{a}^2+k}{a^2}\right)^2-\\
&-&\frac{1}{M^2}(\frac{\dot{a}}{a^2}\frac{d^3a}{dt^3}-\frac{\ddot{a}^2}{a^2}+2\frac{\ddot{a}\dot{a}^2}{a^3}-3(\frac{\dot{a}}{a})^4-2k\frac{\dot{a}^2}{a^4}+\frac{k^2}{a^4}),\nonumber
\eea
where $H^2=\frac{\pi}{8Gk_1}$, $M^2=-\frac{\pi}{8Gk_2}$, with $k_2<0$,
effectively $H$ is the Hubble constant. In this case one has the very simple form for the Ricci tensor: $R^a_b=-3H^2\delta^a_b$.

The solution of (\ref{sta1}) was explicitly obtained in \cite{Staro} where it was found that the de Sitter-like solution is possible, with the scale factor given by $a(t)=H^{-1}\cosh Ht$, or $a(t)=a_0\exp Ht$, or $a(t)=H^{-1}\sinh Ht$, for closed, flat and open Universe respectively. So, we see that accelerating solution is possible in this theory, just as in the presence of the cosmological term. Moreover, it is clear that a wide class of models involving higher orders in curvatures will admit accelerated solutions as well.
This result called interest to $f(R)$ gravity displaying it to be a possible candidate for a consistent explanation of cosmic acceleration. Afterwards, many cosmological solutions for various versions of the function $f(R)$ were obtained and observationally tested, some of these results will be discussed in the next section.

Now, let us discuss the problem of degrees of freedom in $R^2$-gravity. First of all, we note that there is a common difficulty characteristic for higher-derivative theories, either gravitational or not. Indeed, in any Lorentz-invariant theory with four derivatives, the propagator will be proportional to the momentum depending factor looking like:
\bea
\label{propprod}
f(k)=\frac{1}{k^2-\frac{k^4}{M^2}},
\eea
where $M^2$ is the energy scale at which the higher derivatives become important.
It is clear that we can rewrite this factor as
\bea
f(k)=\frac{1}{k^2}-\frac{1}{k^2-M^2}.
\eea
Therefore we see that this propagator actually describes two distinct degrees of freedom, the massive and the massless one. Moreover, these two contributions to the propagator have opposite signs (otherwise, if signs of these contributions are the same, the UV behavior of the propagator is not improved). Clearly it means that the Hamiltonian describing these two degrees of freedom is composed by two terms with opposite signs:
\bea
{\cal H}=\frac{1}{2}(\pi_1^2+\pa_i\phi_1\pa_i\phi_1)-\frac{1}{2}(\pi_2^2+\pa_i\phi_2\pa_i\phi_2+M^2\phi^2_2).
\eea
We see that the energy is not bounded from below, hence, we cannot define a vacuum in the theory consistently, i.e. one can take energy from the system without any limitations, as from a well without a bottom. Moreover, actually it means that the spectrum of the theory describes free particles with negative energy which seems to be nonsense from the viewpoint of the common sense. Actually this is the simplest example of the so-called Ostrogradsky instability plaguing higher-derivative field theory models except of special cases, see a detailed discussion of this example and similar situations in \cite{HH}; a profound discussion of difficulties arising within the Hamiltonian formulation of these theories is given also in \cite{Woodard}. Moreover, in some cases the higher-derivative theories involve not only ghosts but even tachyons, for a specific sign of the higher-derivative term. Therefore the higher-derivative models including the $R^2$-gravity are treated as effective theories aimed for description of the low-energy dynamics of the theory (roughly speaking, for the square of momentum much less than the characteristic mass $M^2$). However, it is necessary to note that higher-derivative terms naturally emerge as quantum corrections after the integration over some matter fields, see f.e. \cite{AM}, so, the presence of higher-derivative terms within the effective dynamics in many field theory models including gravity is natural.

Within our $R^2$-gravity model, the presence of ghosts can be illustrated as follows. If one will extract only physical degrees of freedom, whose role is played by transverse-traceless parts of spatial components $h_{ij}=K_{ij}+F_{ij}$ of the metric fluctuation, and scalar fields, one will see that the quadratic action will look like \cite{Stelle1}
\bea
\label{linact}
{\cal L}_K&=&-\frac{\gamma}{4}K_{ij}\Box K_{ij}+\frac{\gamma}{4}F_{ij}(\Box+m^2_2)F_{ij}-\nonumber\\ &-&
\frac{1}{8}h^T[(8\beta-3\alpha)\kappa^2\Box+\gamma]\Box h^T+\ldots,
\eea
where $h_T=h_{ii}-\nabla^{-2}h_{ij,ij}$ is a trace part. We see that here, $K_{ij}$ and $F_{ij}$ behave as two degrees of freedom, with one of them is massive and another is massless, and their signs are opposite. Hence, the ghost contributions emerge naturally. We see that the number of degrees of freedom is increased, besides of tensor modes we have also scalar ones, and each of them is contributed by usual and ghost ones (the contribution for the scalar $h^T$ can be also split into usual and ghost parts).

Clearly, the natural question is -- whether is it possible to deal with ghosts or even avoid their presence? There are several answers to this question. One approach is based on extracting so-called "benign" ghosts whose contribution can be controlled \cite{Smilga}. Another approach is based on considering the theory where the propagator has a form of the primitive monomial rather than the product of monomials as in (\ref{propprod}). The simplest manner to do it consists in treating of the Lagrangian involving only higher-derivative term with no usual two-derivative one. Within the gravity context it means that one introduces the so-called pure $R^2$ gravity where the usual Einstein-Hilbert term is absent. This theory was introduced in \cite{Alv}, with its action can be treated as the special limit of $R^2$ gravity: $S=\sqrt{|g|}(\beta R^2+\kappa^{-2}R)$, with $\kappa^{-2}\to 0$. The propagator will be proportional to
\bea
G_{\mu\nu\rho\sigma}(k)=\frac{1}{6\beta}\frac{1}{k^4}P^0_{\mu\nu,\rho\sigma},
\eea
with $P^0_{\mu\nu,\rho\sigma}=\frac{1}{3}P_{\mu\nu}P_{\rho\sigma}$, the $P_{\rho\sigma}$ is the usual transverse projector, and $\beta$ is a coefficient at $R^2$. One can show that on the flat background, only scalar mode propagates \cite{Alv}. It is clear that there is no ghosts in this theory (in \cite{Alv} it is also argued with analysis of degrees of freedom). It is interesting to note that the Breit potential for this propagator displays confining behavior:
\bea
V(\vec{r})=\int\frac{d^3k}{(2\pi)^4}\frac{e^{i\vec{k}\cdot \vec{r}}}{\vec{k}^4}\propto |\vec{r}|.
\eea 
So, this theory has only one difficulty -- it does not yield Einstein-Hilbert limit which was tested through many observations. Many aspects of the pure $R^2$-gravity are discussed in \cite{Salvio}, see also references therein. 

\section{$f(R)$-gravity}

Clearly, the natural development of the idea of $R^2$ gravity will consist in the suggestion that the classical action can involve not only second but any degree (involving negative!) of the scalar curvature. Thus, the concept of $f(R)$ gravity was introduced. Its action is given by (\ref{fr}), with $f(R)=R+\gamma R^n$.

First of all, we can discuss the renormalizability of this theory along the same lines as in the previous section. It is easy to see that the term proportional to $R^N$ (or, which is similar, to $N$-th degree or Riemann or Ricci tensors) is characterized by the degree of divergence $\omega$, in the four-dimensional space-time given by
\bea
\omega=4L-2n(P-V)-2N=(4-2n)L+2n-2N.
\eea
Immediately we see that now discussion of the renormalizability is more involved than for $n=2$ (the similar situation occurs for Horava-Lifshitz-like theories where increasing of the critical exponent $z$ implies in growing not only of degree of momentum in the denominator of the propagator but also of numbers of derivatives in vertices). Actually, for any $n>2$ one should classify possible divergences with various values of $N$ for the given $n$. Many examples of quantum calculations in theories for various $n$, as well as in other higher-derivative gravity theories, including studies of one-loop divergences and running couplings are presented in \cite{BOS}, see also references therein. It is clear that the ghosts will arise for any polynomial form of $f(R)$ just as in the case of $R^2$-gravity, so, conceptually the quantum calculations for $n=2$ and for $n>2$ do not differ essentially (for discussion of renormalizability aspects of $f(R)$ gravity, see also \cite{ShaSalles}).

The main line of study of $f(R)$ gravity consists in a detailed investigation of its classical, especially cosmological aspects. The modified Einstein equations in this case look like
\bea
\label{genmov}
f^{\prime}(R)R_{\mu\nu}-\frac{1}{2}g_{\mu\nu}f(R)+(g_{\mu\nu}\nabla_{\lambda}\nabla^{\lambda}-\nabla_{\mu}\nabla_{\nu})f^{\prime}(R)=8\pi G T_{\mu\nu}.
\eea
It is evident that the dS/adS spaces will be vacuum solutions of these equations yielding $f(R)=bR^2+\Lambda$, with $b$ being a constant. Then, to study the cosmological aspects, we can use the expressions for components of the Ricci tensor and the scalar curvature (\ref{cosmcurv}). In a whole analogy with (\ref{sta1}) one can find that, if the $f(R)$ involves $R^2$ term, the corresponding cosmological equation will be
\bea
\label{sta2}
\frac{\dot{a}^2+k}{a^2}&=&\frac{1}{H^2}\left(\frac{\dot{a}^2+k}{a^2}\right)^2-
\frac{1}{M^{2n}}\left(\frac{\dot{a}^{2n}}{a^{2n}}+\ldots\right),
\eea
where $H$ is the constant, accompanying the $R^2$ term, cf. (\ref{sta1}), and $M$ is the constant accompanying the higher curvature term. The dots in parentheses are for other terms with $2n$ time derivatives (if $k=0$ they are all homogeneous, involving the same degrees of $a$ in the numerator and in the denominator). It can be shown (see f.e. \cite{Woodard} and references therein), that in this theory, for any $n\geq 2$ the solutions are again presented by hyperbolic sine and cosine and exponential, just as in $R^2$ case \cite{Staro}. 
We conclude that this theory describes well the inflationary epoch where the curvature of the Universe was large hence the higher-derivative contributions are important. In principle, in this earlier epoch one can use the action introduced in the manner of \cite{Alv,Salvio} where the Einstein-Hilbert term is suppressed, and one chooses $f(R)=\gamma R^n$ as a reasonable approximation.
At the same time, an interesting problem is -- how one can adopt the form of the $f(R)$ to explain the actual accelerated expansion of the Universe, in the case where the curvature is very close to zero, so,  $R^n$ terms with $n>1$ can be disregarded. 

In \cite{CarTro}, a bold departure from usual forms of the $f(R)$ function was proposed: this function was suggested to be
\bea
\label{1r}
f(R)=R-\frac{\mu^4}{R}.
\eea 
The quantum description of this theory near the flat background is problematic. However, it can be treated perturbatively in principle near some other background. 

Let us discuss the equations of motion for this choice of $f(R)$. In the vacuum case ($T_{\mu\nu}=0$), we have
\bea
\label{mov1}
(1+\frac{\mu^4}{R^2})R_{\mu\nu}-\frac{1}{2}(1-\frac{\mu^4}{R^2})Rg_{\mu\nu}+(g_{\mu\nu}\Box-\nabla_{\mu}\nabla_{\nu})\frac{\mu^4}{R^2}=0.
\eea 
For the constant scalar curvature, one finds
\bea
R_{\mu\nu}=\pm\frac{\sqrt{3}}{4}\mu^2g_{\mu\nu},
\eea
this is (a)dS solution, and in the case of the negative sign, at $\mu\neq 0$ we indeed have an acceleration \cite{Woodard}, so, this model allows to explain accelerated expansion for the constant curvature case. 

Unfortunately, this model suffers from a tachyonic instability. Indeed, after taking the trace of (\ref{mov1}) we find
\bea
-R+\frac{3\mu^4}{R}+3\Box(\frac{\mu^4}{R^2})=0.
\eea
After we make a perturbation $\delta R$ around an accelerated solution described by a constant negative curvature, i.e. $R=-\sqrt{3}\mu^2+\delta R$, we find that the $\delta R$ obeys the equation
\bea
-\delta R+\frac{2}{\sqrt{3}\mu^2}\Box\delta R=0,
\eea
and in our signature $(+---)$ this equation describes a tachyon. Actually, this instability is very weak since $\mu^2$ is observationally very small, hence the first term in this equation is highly suppressed. It should be noted that for a non-zero density of the matter the instability is much worse, but adding the $R^2$ term into the action improves radically the situation \cite{Woodard}. Therefore this model was naturally treated as one of candidates for solving the dark energy problem. However, the model (\ref{1r}), in further works, was discussed mostly within the cosmological context (see also a discussion of asymptotic behavior of cosmological solutions in \cite{CarTro}).

Let us note some more issues related to $f(R)$ gravity. First, it was argued in \cite{Woodard} that the $f(R)$
 gravity model is equivalent to a some scalar-tensor gravity. Indeed, let us for the first step define $f(R)=R+\bar{f}(R)$, so $\bar{f}(R)$ is a correcting term. Then, we introduce an auxiliary scalar field $\phi=1+\bar{f}^{\prime}(R)$. Since this equation relates $R$ and $\phi$, it can be solved, so one obtains a dependence $R=R(\phi)$. As a next step, the potential looking like
\bea
U(\phi)=(\phi-1)R(\phi)-\bar{f}(R(\phi)),
\eea 
implying $U^{\prime}(\phi)=R(\phi)$, is defined. As a result, the Lagrangian (\ref{fr}) turns out to be equivalent to
\bea
{\cal L}_E=\sqrt{|g|}(\phi R-U(\phi)).
\eea
Then, we carry out the conformal transformation of the metric:
\bea
\tilde{g}_{\alpha\beta}=\phi g_{\alpha\beta},\quad\, \phi=\exp(\sqrt{\frac{4\pi G}{3}}\varphi),
\eea
therefore the Lagrangian is rewritten as
\bea
{\cal L}_E&=&\sqrt{|\tilde{g}|}(\frac{1}{16\pi G}\tilde{R}-\frac{1}{2}\tilde{g}^{ab}\pa_a\varphi\pa_b\varphi-V(\varphi)),\nonumber\\
V(\varphi)&=&\frac{1}{16\pi G}U\left(\exp(\sqrt{\frac{4\pi G}{3}}\varphi)\right)\exp\left(-\sqrt{\frac{16\pi G}{3}}\varphi\right).
\eea
Therefore, the $f(R)$ gravity turns out to be equivalent to the general relativity with the extra scalar, i.e. to the scalar-tensor gravity. The form of the potential is therefore related with the form of the function $f(R)$.
  
Clearly, the natural question is about possibility to obtain other important gravitational solutions  within the $f(R)$ gravity context.  First, for the G\"{o}del metric (\ref{Godel}), as well as for its straightforward generalization defined in \cite{RebSan} as G\"{o}del-type metric:
\begin{equation}
\label{gtype}
ds^2=(dt+H(r)d\phi)^2-D^2(r)d\phi^2-dr^2-dz^2,
\end{equation}
where
\begin{equation}
\label{condhom}
\frac{H^{\prime}}{D}=2\omega,\quad\, \frac{D^{\prime\prime}}{D}=m^2,
\end{equation}
with $\omega,m$ are constants, 
 the scalar curvature is constant, hence the equations (\ref{genmov}) are simplified drastically since the term involving covariant derivatives of $f(R)$ goes away, and the l.h.s. of these equations turns out to be a mere combination of constants. It was shown in \cite{RebSan} that both causal and non-causal solutions are possible, with $f(R)$ is an arbitrary function of the scalar curvature, while to achieve causality, it is not sufficient to have only a relativistic fluid as in \cite{Godel}, and one must add as well a scalar matter -- one should remind that since the Einstein equations are nonlinear, the solution generated by a sum of two sources is  not equal to the sum of solutions generated by each source. As for the black holes, we strongly recommend the excellent book \cite{Lobo} where Schwarzschild-type BH solutions in $f(R)$ gravity are considered, see also \cite{BH1} and references therein. 
 
In \cite{Lobo}, a wide spectrum of possible generalizations of $f(R)$ gravity was discussed, such as $f(R,{\cal L}_m)$ and $f(R,T)$ models, where ${\cal L}_m$ is the matter Lagrangian, and $T$ is the trace of the energy-momentum tensor. However, within our study we will pursue another aim -- we will suggest that the matter is coupled to the gravity in the usual form while the free gravity action depends on other scalars constructed on the base of the Riemann tensor and metric. This will be the subject of the next section.

\section{Functions of other curvature invariants}

Let us suggest that instead of the function of the scalar curvature only, we have also functions of other scalars. There are many examples of studies of such models, so we discuss only some most interesting ones, the $f(R,Q)$ gravity, the Lovelock gravity and the Gauss-Bonnet gravity.

We start our discussion from the $f(R,Q)$ gravity. In this theory, the Lagrangian is a function not only of the scalar curvature, but also of $Q=R_{\mu\nu}R^{\mu\nu}$, so,
\bea
S=\int d^4x \sqrt{|g|}f(R,Q)+S_m.
\eea
  The equations of motion are found to look like \cite{Olmo1,ourfrq}
\begin{eqnarray}
& & f_{R}R_{\mu\nu}-\frac{f}{2}g_{\mu\nu}+2f_{Q}R_{(\mu}^{\beta}R_{\nu)\beta}+g_{\mu\nu}\Box f_{R}-\nabla_{(\mu}\nabla_{\nu)}f_{R}+\nonumber\\
&+&\Box\big(f_{Q}R_{\mu\nu}\big)-2\nabla_{\lambda}\big[\nabla_{(\mu}\big(f_{Q}R^{\lambda}_{\nu)}\big)\big]+g_{\mu\nu}\nabla_{\alpha}\nabla_{\sigma}\big(f_{Q}R^{\alpha\sigma}\big)=\kappa^2 T_{\mu\nu}^m,
\label{FE1}
\end{eqnarray}    
where $f_Q=\frac{\partial f}{\partial Q}$, $f_R=\frac{\partial f}{\partial R}$, and $T_{\mu\nu}^m$ is the energy-momentum tensor of the matter.

As an example, we consider the G\"{o}del-type metric (\ref{gtype}). One can show, that, unlike general relativity, such solutions are possible not only for dust but also for the vacuum (with non-zero cosmological constant), in particular, completely causal vacuum solutions are present \cite{ourfrq}. Clearly, the solutions of this form are possible also for the presence of the matter given by the relativistic fluid and a scalar field. Again, as in \cite{RebSan}, all Einstein equations will take the form of purely algebraic relations between density, pressure, field amplitude and constants from the gravity Lagrangian.
As for the cosmological metric, the possibility of accelerating solutions can be shown just in the same manner as in the previous sections. Among other possible solutions in $f(R,Q)$ gravity, it is worth to mention Reissner-Nordstr\"{o}m black holes \cite{Olmo2} and wormholes \cite{Olmo3}. Further generalization of this theory would consist in consideration of function not only of $R$ and $Q$, but also of $P=R_{\mu\nu\alpha\beta}R^{\mu\nu\alpha\beta}$, with study of the corresponding theory called $f(R,Q,P)$ gravity is in principle not more difficult, see f.e. \cite{Cognola1}. 

Now, let us make the next step -- suggest that the dimension of the space-time is not restricted to be four but can be arbitrary. This step allows us to introduce the Lovelock gravity. Its key idea is as follows.

Let us consider the gravity model defined in the space-time of an arbitrary dimension \cite{Lovelock}, called the Lovelock gravity:
\bea
\label{lovelock}
S=\int d^D x \sqrt{|g|}(c_0\Lambda+c_1R+c_2{\cal G}+\ldots).
\eea
Here $c_0,c_1,c_2,\ldots$ are some constants possessing nontrivial dimensions. It is natural to suggest that they, up to some dimensionless numbers, are given by various degrees of the gravitational constant.
Each term with $2n$ derivatives is topological, i.e. it represents itself as a total derivative at $D=2n$, and identical zero in minor dimensions. We note that there is no higher derivatives of the metric in the action. This action is characterized the following properties displayed by the Einstein-Hilbert action: (i) the tensor $A_{\alpha\beta}$, the l.h.s. of the corresponding equations of motion, is symmetric; (ii) the covariant divergence of $A_{\alpha\beta}$ vanishes; (iii) the $A_{\alpha\beta}$ is linear in second derivatives of the metric. 

The general form of the term with $2n$ derivatives in the Lagrangian contributing to (\ref{lovelock}) can be presented as \cite{Deruelle}:
\bea
{\cal L}_n=\frac{1}{2^n}\delta^{i_1\ldots i_{2n}}_{j_1\ldots j_{2n}}R^{j_1j_2}_{\phantom{j_1j_2}i_1i_2}\ldots 
R^{j_{2n-1}j_{2n}}_{\phantom{j_{2n-1}j_{2n}}i_{2n-1}i_{2n}},
\eea
where the $2n$-order Kronecker-like delta symbol is
\bea
\delta^{i_1\ldots i_{2n}}_{j_1\ldots j_{2n}}=\left|\begin{array}{ccc}
\delta^{i_1}_{j_1} &\ldots & \delta^{i_1}_{j_{2n}}\\
\ldots & \ldots & \ldots \\
\delta^{i_{2n}}_{j_1} & \ldots & \delta^{i_{2n}}_{j_{2n}}
\end{array}
\right|.
\eea
It is easy to check that at $n=1$, we have the scalar curvature, and at $n=2$, the Gauss-Bonnet term. The term with $n=0$ is naturally treated as the cosmological constant. 
As a result, we can write down the action;
\bea
S=\frac{1}{\kappa^2}\int d^Dx\sqrt{|g|}\sum_{0\leq n < D/2}\alpha_n \lambda^{2(n-1)}{\cal L}_n.
\eea
Here, zero order is for $\Lambda$, first -- for $R$, second -- for ${\cal G}$. The $\alpha_n$ are some numbers, and $\lambda$ is a length scale, f.e. Planck length, it is given by $\kappa$ in $D=4$ where the $\kappa^{-1}$ has a dimension of inverse length.

The l.h.s. of the modified Einstein equations looks like \cite{Deruelle}
\bea
\label{lhs}
G_{\alpha\beta}&=&\sum_{0\leq n \leq D/2}\alpha_n \lambda^{2(n-1)}G_{(n)\alpha\beta};\nonumber\\
G_{(n)\phantom{\alpha}\beta}^{\phantom{(n)}A}&=&-\frac{1}{2^{n+1}}\delta^{\alpha i_1\ldots i_{2n}}_{\beta j_1\ldots j_{2n}}
R^{j_1j_2}_{\phantom{j_1j_2}i_1i_2}\ldots 
R^{j_{2n-1}j_{2n}}_{\phantom{j_{2n-1}j_{2n}}i_{2n-1}i_{2n}}.
\eea
It is clear that $G_{(0)\alpha\beta}=-\frac{1}{2}g_{\alpha\beta}$, $G_{(1)\alpha\beta}=G^{EH}_{\alpha\beta}\equiv R_{\alpha\beta}-\frac{1}{2}Rg_{\alpha\beta}$ is the usual Einstein tensor.
The r.h.s. of the modified Einstein equations is not modified within this approach, so we have $G_{\alpha\beta}=\kappa^2 T_{\alpha\beta}$.

It turns out to be that although the l.h.s. (\ref{lhs}) of the modified Einstein equations is very complicated, these equations admit some exact solutions for an arbitrary space-time dimension, i.e. for the presence of terms with very high orders in curvatures. The most interesting cases are the maximally symmetric (anti) de Sitter space and the FRW cosmological metric.
 
In the (a)dS space, the Riemann curvature tensor is given by
\bea
R_{\alpha\beta\gamma\delta}=\frac{\sigma}{\lambda^2} (g_{\alpha\gamma}g_{\beta\delta}-g_{\alpha\delta}g_{\beta\gamma}),
\eea
with $\sigma$ is a some number. In this case the vacuum equation yields $\sum\limits_{0\leq n < D/2}\beta_n\sigma^n=0$, with $\beta_n=\frac{(D-1)!}{(D-2n-1)!}\alpha_n$, and this equation possesses some roots for $\sigma$ (in general complex ones). Each value of $\sigma$ allows to find the corresponding scalar curvature.
 
 We can solve the modified Einstein equations also for the FRW metric (\ref{FRW}):
 \bea
R_{i0j0}=-g_{ij}\frac{\ddot{a}}{a};\quad\, R_{ijkl}=\frac{\dot{a}^2+k}{a^2}(g_{ik}g_{jl}-g_{il}g_{jk}),
\eea
with $A=\frac{\lambda^2(k+\dot{a}^2)}{a^2}$, where as usual $k=-1,0,1$, the indices $i,j,k,l$ take values 1,2,3, and the l.h.s. of modified Einstein equations yields
\bea
\lambda^2G_{00}&=&\frac{1}{2}\sum_{0\leq n <D/2}\beta_nA^n,\\
\lambda^2G_{ij}&=&-\frac{1}{2}\frac{g_{ij}}{2(D-1)}\sum_{0\leq n < D/2}\beta_nA^{n-1}[2n\lambda^2\frac{\ddot{a}}{a}+(D-2n-1)A].\nonumber
\eea
For the vacuum one immediately finds $A=const$ which implies exponential expansion. For the fluid, also there are hyperbolic and trigonometric solutions. So, we conclude that the Lovelock gravity is consistent with accelerating expansion of the Universe.

Concerning the general Lovelock theory, it must be noted that already third-order contributions to the action, those ones with six derivatives, imply very complicated equations of motion. The explicit expressions for initial terms of Lovelock Lagrangians up to fifteenth order (for which, the whole expression involves tens of millions of terms) can be found in \cite{Briggs}.
 
Now, let us discuss the Gauss-Bonnet gravity in the arbitrary spacetime dimension, that is, the theory with the action
\bea
S=\int d^Dx\sqrt{|g|}(\frac{1}{2\kappa^2}R+f({\cal G})+{\cal L}_m).
\eea
The equations of motion of this theory are
\bea
\label{eqmod}
\frac{1}{\kappa^2}G_{\mu\nu}&=&2T_{\mu\nu}+\frac{1}{2}g_{\mu\nu}f({\cal G})-2F({\cal G})RR_{\mu\nu}+4F({\cal G})R_{\mu}^{\lambda}R_{\nu\lambda}-\nonumber\\ 
&-& 2F({\cal G})R_{\mu\lambda\rho\sigma}R_{\nu}^{\phantom{\nu}\lambda\rho\sigma}-
4F({\cal G})R_{\mu\rho\sigma\nu}R^{\rho\sigma}+2R\nabla_{\mu}\nabla_{\nu}F({\cal G})-\nonumber\\&-&
2Rg_{\mu\nu}\nabla^2F({\cal G})-4R_{\mu}^{\rho}\nabla_{\nu}\nabla_{\rho}F({\cal G})\nonumber\\&-& 4R_{\nu}^{\rho}\nabla_{\mu}\nabla_{\rho}F({\cal G})+4R_{\mu\nu}\nabla^2F({\cal G})+4g_{\mu\nu}R^{\lambda\rho}\nabla_{\lambda}\nabla_{\rho}F({\cal G})-\nonumber\\
&-&4R_{\mu\nu\lambda\rho}\nabla^{\lambda}\nabla^{\rho}F({\cal G})\nonumber\\
&\equiv& 2T_{\mu\nu}+H_{\mu\nu},
\eea
with $F({\cal G})=f'({\cal G})$. As a simple example, we discuss the solution of this equation in the braneworld case, i.e. we consider the five-dimensional metric
\bea
ds^2=g_{\mu\nu}dx^{\mu}dx^{\nu}=e^{2A(y)}\eta_{ab}dx^adx^b-dy^2,
\eea
where we suggest that the indices $\mu,\nu$ vary from 0 to 4 while $a,b$ -- from 0 to 3, and $y$ is the extra (fourth) spacial coordinate, and $A(y)$ is called the warp factor \cite{RS}. In \cite{ourGB}, these equations have been solved for the case when the matter is given by the scalar field $\phi$, so, $T_{ab}=\eta_{ab}e^{2A}(\frac{1}{2}\phi^{\prime 2}+V(\phi))$, and $T_{44}=\frac{1}{2}\phi^{\prime 2}-V(\phi)$, for the simplified situation ${\cal G}=const=\pm 120b^2$. Explicitly, 
for positive $B(={\cal G}/120)=b^2$, the solution is 
\bea
y+C=\frac{4}{5}\int\frac{dA^{\prime}(A^{\prime})^2}{b^2-(A^{\prime})^4},
\eea
and for the negative $B=-b^2$ -- in the form
\bea
y+C=-\frac{4}{5}\int\frac{dA^{\prime}(A^{\prime})^2}{b^2+(A^{\prime})^4}.
\eea
In principle, there are more situations when the modified Einstein equations in the Gauss-Bonnet gravity can be solved. We note that the braneworld solutions could be found not only for Gauss-Bonnet gravity but for other gravity models including the already discussed $f(R)$ gravity (see f.e. \cite{BLMP} and references therein), however, we do not discuss the details of braneworld solutions here because of the restricted volume of this review.

\section{Conclusions}

We discussed various extensions of Einstein gravity characterized by modifications in the purely gravitational sector. These modifications are based on  adding new scalars representing themselves not only as various degrees of the scalar curvature but also as functions of higher order curvature invariants. We explicitly demonstrated that the $R^2$ gravity is all-loop renormalizable, and that the most important solutions of general relativity, such as cosmological FRW metric and G\"{o}del metric continue to be solutions within modified gravity. Moreover, we showed that modifications of the pure gravitational sector allow for accelerated cosmological expansion being thus examples of reasonable solutions for the dark energy problem, so that the problem of choosing a better modification of the gravity apparently can be solved in principle while the problem of choice for the most adequate modification of the gravity is actually more observational and experimental than theoretical.

Within this section we presented several other interesting results. First, we described the argumentation allowing for establishing the equivalence between modifications in the pure gravitational sector and adjusting the action of the extra scalar field coupled to gravity, which implies that the $f(R)$ gravity is equivalent to the scalar-tensor gravity with an appropriate potential. Second, we discussed the $1/R$ terms whose form seems to be highly controversial since the observed curvature of the space-time is very small hence these terms are very large. Third, we considered possible generalizations of the gravity consistent within the extra dimensions concept.

Further development of a general gravity model consists in the idea that for the complete description of the gravity it is not sufficient to study only the metric, so that the gravity model should be extended through adding some other fields which are necessary being fundamental ingredients of the complete theory. So, one must consider scalar-tensor and vector-tensor gravity models. We will consider  some examples of these models in our next chapters.

\chapter{Scalar-tensor gravities}

\section{General review}

In the previous chapter we demonstrated that modifications of the pure gravitational sector allow for obtaining interesting results, in particular,  for a consistent explanation of the cosmic acceleration. At the same time, we noted that $f(R)$ gravities are dynamically equivalent to some gravity models whose action is given by the sum of the usual Einstein term and a new term depending on the extra scalar field \cite{Woodard}. This field, being related with the function of the curvature, evidently cannot be associated with the matter, hence it is natural to suggest that the complete description of gravity is given by composition of the dynamical metric tensor and this scalar field, so we have the scalar-tensor  gravity model. Another motivation for a scalar-tensor gravity arises from quintessence models in cosmology which involve a very light scalar field called the quintessence field and are known to explain accelerated expansion of the Universe as well as the cosmological constant which therefore implied active application of the quintessence field within the inflationary context \cite{RatraPeebles}. The advantage of the quintessence in comparison with the cosmological constant consists in the fact that the very tiny mass of the quintessence field (estimated to be about $10^{-33}$ eV \cite{Tsuji}) is much more reasonable from the theoretical viewpoint than the extremal smallness of the cosmological constant giving the famous cosmological constant problem, since even the massless scalar fields are physically consistent.

While the quintessence is well discussed now (see f.e. \cite{Tsuji} and references therein), there are other interesting manners to introduce new scalar fields in the gravity, moreover, while the quintessence field is treated as a matter, the scalar fields introduced within these approaches are interpreted as ingredients of the complete description of the gravity rather than the matter. One of these manners is the Brans-Dicke gravity where the gravitational constant whose negative dimension is responsible for a non-renormalizability of the gravity is suggested to be not a fundamental constant but a function of a some slowly varying fundamental scalar field. Another one is the four-dimensional Chern-Simons modified gravity where the pseudoscalar field allows to implement the CPT (and in certain cases Lorentz) symmetry breaking in the gravity context. And actually, one more model is intensively discussed in this context, that is the galileons model. Namely these theories will be considered in this chapter.

\section{Chern-Simons modified gravity}

\subsection{The $4D$ Chern-Simons modified gravity action}

The three-dimensional Chern-Simons (CS) term has been originally introduced in the paper \cite{DJT} within the context of electrodynamics, as an example of a term conciliating gauge invariance with a non-zero mass. It has been immediately generalized to the non-Abelian case, so, the CS Lagrangian looks like
\bea
\label{cs}
{\cal L}^{A}_{CS}=\epsilon^{\mu\nu\lambda}(A_{\mu}^a\partial_{\nu}A_{\lambda}^a+\frac{2}{3}f^{abc}A_{\mu}^a A_{\nu}^bA_{\lambda}^c),
\eea
where $A_{\mu}=A_{\mu}^aT^a$ is the Lie-algebra valued gauge field, and $f^{abc}$ are the structure constants. In the gravity case, the role of the gauge field is played by the connection, and the three-dimensional gravitational CS term reads as \cite{DJT,JaPi}:
\begin{equation}
\label{fullcs}
S_{CS}=\frac{1}{2\kappa^2 \mu}\int d^3x \epsilon^{\mu\nu\lambda}(\Gamma_{\mu a}^{\phantom{a}b}\partial_{\nu}\Gamma_{\lambda b}^{\phantom{b}a}+\frac{2}{3}\Gamma_{\mu a}^{\phantom{a}b}\Gamma_{\nu b}^{\phantom{b}c}\Gamma_{\lambda c}^{\phantom{c}a}).
\end{equation}
In principle, in non-Riemannian geometries we can use an independent connection rather than the Levi-Civita one, however, this general situation is outside of the scope of our review. Here, the $\epsilon^{\mu\nu\lambda}$, which can take values $1,0,-1$, is the usual Levi-Civita symbol, not the covariant one. Varying the CS term with respect to the metric, one finds
\bea
\delta S_{CS}=-\frac{1}{\kappa^2\mu}\int d^3x C^{\mu\nu}\delta g_{\mu\nu},
\eea
where
\bea
C^{\mu\nu}=-\frac{1}{2\sqrt{|g|}}\epsilon^{\mu\alpha\beta}\nabla_{\alpha}R^{\nu}_{\beta}+(\mu\leftrightarrow \nu)
\eea
is the three-dimensional Cotton tensor. It is evidently symmetric and traceless. The $\mu$ is a some constant of the mass dimension 1. So, the modified Einstein equations look like
\bea
G^{\mu\nu}+\frac{1}{\mu}C^{\mu\nu}=0.
\eea
It is useful also to write the linearized form of the gravitational Chern-Simons action obtained from (\ref{fullcs}) under the replacement $g_{\mu\nu}=\eta_{\mu\nu}+\kappa h_{\mu\nu}$:
\bea
\label{fullCS0}
S^{(0)}=-\frac{1}{2\mu}\int d^3x h^{\mu\nu}\epsilon_{\alpha\mu\rho}\partial^{\rho}(\Box \eta_{\gamma\nu}-\partial_{\gamma}\partial_{\nu})h^{\gamma\alpha}.
\eea
We see that this action is, first, explicitly gauge invariant under usual linearized gauge transformations $\delta h_{\mu\nu}=\pa_{\mu}\xi_{\nu}+\pa_{\nu}\xi_{\mu}$, second, involves higher derivatives. However, after obtaining the equations of motion for the full linearized action formed by the sum of the terms (\ref{FP}) and (\ref{fullCS0}), one finds that the physical degrees of freedom satisfy the second-order equation \cite{DJT}, with their propagator behaves as $(\Box+\mu^2)^{-1}$, thus, in the $3D$ CS modified gravity there is no problems with negative-energy states discussed in the previous chapter. The similar situation occurs in the four-dimensional case as well.

The generalization of this theory to the four-dimensional case turns out to be straightforward, however, in this case, similarly to the electrodynamics, this generalization essentially involves the CPT (and in certain cases Lorentz) symmetry breaking. From the formal viewpoint such a generalization for the linearized theory is performed through replacement $\epsilon^{\mu\nu\lambda}\to b_{\rho}\epsilon^{\rho\mu\nu\lambda}$, with $b_{\rho}$ is a constant vector, which allows to convert the CS term to the Carroll-Field-Jackiw (CFJ) term which in the Abelian case looks like
\bea
\label{CFJ}
{\cal L}_{CFJ}=\epsilon^{\rho\mu\nu\lambda}b_{\rho} A_{\mu}\partial_{\nu}A_{\lambda}.
\eea
In principle, such a replacement of the three-dimensional Levi-Civita symbol by the four-dimensional one contracted with a vector already allows to write down the four-dimensional gravitational CS term:
\bea
\label{fullCS03}
{\cal L}_{CS,grav}=\int d^4x  \epsilon^{\rho\mu\nu\lambda}b_{\rho}(\Gamma_{\mu a}^{\phantom{a}b}\partial_{\nu}\Gamma_{\lambda b}^{\phantom{b}a}+\frac{2}{3}\Gamma_{\mu a}^{\phantom{a}b}\Gamma_{\nu b}^{\phantom{b}c}\Gamma_{\lambda c}^{\phantom{c}a}),
\eea
with its linearized form is
\bea
\label{fullCS04}
S^{(0)}=-\frac{1}{2}\int d^4x h^{\mu\nu}\epsilon_{\alpha\mu\rho\lambda}b^{\lambda}\partial^{\rho}(\Box \eta_{\gamma\nu}-\partial_{\gamma}\partial_{\nu})h^{\gamma\alpha}.
\eea
We note that this action is invariant under the same linearized gauge transformations $\delta h_{\mu\nu}=\pa_{\mu}\xi_{\nu}+\pa_{\nu}\xi_{\mu}$.
Now, it is very interesting to discuss some motivations for this term.
  
First of all, already in 1984, much time before the interest to Lorentz-CPT breaking strongly increased, the gravitational anomalies have been discussed in \cite{AlvAnom}, where the topological current $K^{\mu}$ was introduced, with its explicit form is
\bea
K^{\rho}=2\epsilon^{\rho\mu\nu\lambda}(\Gamma_{\mu a}^{\phantom{a}b}\partial_{\nu}\Gamma_{\lambda b}^{\phantom{b}a}+\frac{2}{3}\Gamma_{\mu a}^{\phantom{a}b}\Gamma_{\nu b}^{\phantom{b}c}\Gamma_{\lambda c}^{\phantom{c}a}),
\eea 
with its divergence is
\bea
\label{divcurr}
\partial_{\rho}K^{\rho}=\frac{1}{2}\epsilon^{\mu\nu\alpha\beta}R_{\mu\nu\gamma\delta}R_{\alpha\beta}^{\phantom{\alpha\beta}\gamma\delta}\equiv {}^*RR.
\eea
We note that the $3D$ gravitational Chern-Simons term, up to overall multiplier, is equal to the $K^3$ component, i.e. the component of this current directed along "extra", $z$ axis. 

It is clear that the integral from (\ref{divcurr}) over the space-time is a surface term. To include it into the action in a consistent form, one should introduce a new field $\vartheta$ called the CS coefficient. As a result, we can add to the usual Einstein-Hilbert action the new term proportional to $\vartheta$ which we call the CS action $S_{CS}$:
\begin{eqnarray}
\label{csact}
S_{CS}&=&\frac{1}{2\kappa^2}\int d^4x (-\frac{1}{2}v_{\mu}K^{\mu})=\frac{1}{2\kappa^2}I_{CS};\nonumber\\
S_{EH+CS}&=&\frac{1}{2\kappa^2}\int d^4x (\sqrt{-g}R+\frac{1}{4}\vartheta {}^*RR).
\end{eqnarray} 
Here, $v_{\mu}=\partial_{\mu}\vartheta$ is a vector. We note that in principle this vector is rather a function of space-time coordinates than the constant, hence, in general the gravitational CS term breaks the CPT symmetry. However, the $\vartheta$ can be treated as an external, but not dynamical, field, therefore one can choose $v_{\mu}$ to be the constant vector. This immediately implies the Lorentz symmetry breaking, therefore in this case the $4D$ CS modified gravity whose action is given by the second equation in (\ref{csact}) turns out to be the first example of the gravity model with the Lorentz symmetry breaking.

The equations of motion for the CS modified gravity can be easily obtained.  Varying the CS term $I_{CS}$ defined by the first equation in (\ref{csact}), we get
\begin{equation}
\delta I_{CS}=\int d^4x\sqrt{-g}C^{\mu\nu}\delta g_{\mu\nu}, 
\end{equation}
with $\varepsilon^{\alpha\beta\gamma\delta}=\frac{\epsilon^{\alpha\beta\gamma\delta}}{\sqrt{|g|}}$ is a Levi-Civita tensor (not a simple symbol!), and
 \begin{eqnarray}
 C^{\mu\nu}&=&-\frac{1}{2}[v_{\sigma}(\varepsilon^{\sigma\mu\alpha\beta}\nabla_{\alpha}R^{\nu}_{\beta}+
\varepsilon^{\sigma\nu\alpha\beta}\nabla_{\alpha}R^{\mu}_{\beta})+
v_{\sigma\tau}({}^{*}R^{\tau\mu\sigma\nu}+\ ^{*}R^{\tau\nu\sigma\mu})],
\end{eqnarray}   
is the Cotton tensor, and  $v_{\sigma\tau}=\nabla_{\sigma}v_{\tau}$. One can check that the covariant divergence of the Cotton tensor is proportional to the invariant $\ ^{*}RR$:
 \begin{equation}
  \nabla_{\mu}C^{\mu\nu}=\frac{1}{8}v^{\nu}\ ^{*}RR.
\label{nabla}
\end{equation}  
This divergence plays the crucial role when the modified Einstein equations are considered. Their explicit form is
\begin{equation}
\label{mee}
G^{\mu\nu}+C^{\mu\nu}=\kappa^2 T^{\mu\nu},
\end{equation}
so, due to the Eq. (\ref{nabla}), we find that the conservation of the energy-momentum tensor requires the vanishing of the divergence of the Cotton tensor, which, according to (\ref{nabla}), yields an additional consistency condition called the Pontryagin constraint:
\begin{equation}
\label{conscond}
\ ^{*}RR=0,
\end{equation}
 which must be checked for any solution. However, since in many cases, including, among others, the rotational symmetry, the curvature tensor has the structure $R_{[ab][ab]}$, i.e. its only non-zero components are $R_{0101},R_{0202},\ldots$, this consistency condition will be automatically satisfied in these cases.
 
 The further extension of the Chern-Simons modified gravity (CSMG) was carried out through assuming the nontrivial dynamics for the $\vartheta$ CS coefficient. The key idea is as follows \cite{Grumiller}: we assume that the action of CSMG includes the kinetic term for $\vartheta$, looking like
 \begin{eqnarray}
\label{csactdyn}
S=\frac{1}{2\kappa^2}\int d^4x \sqrt{|g|}(R+\frac{1}{2}\nabla^m\vartheta\nabla_m\vartheta-V(\vartheta)-\frac{1}{\alpha}\vartheta {}^*RR),
\end{eqnarray} 
with now ${}^*RR\equiv\frac{1}{2}\varepsilon^{\mu\nu\alpha\beta}R_{\mu\nu\gamma\delta}R_{\alpha\beta}^{\phantom{\alpha\beta}\gamma\delta}$, i.e. it is redefined with the Levi-Civita tensor $\varepsilon^{\mu\nu\lambda\rho}=\frac{\epsilon^{\mu\nu\lambda\rho}}{\sqrt{|g|}}$, and instead of the Pontryagin constraint (\ref{conscond}), one has the equation of motion for $\vartheta$:
\begin{equation}
\label{eqtheta}
{}^*RR=-\alpha(\Box\vartheta+\frac{\partial V}{\partial\vartheta}).
\end{equation}
If we have a metric consistent within the non-dynamical CS framework with a specific $\vartheta$, it is consistent in the dynamical case if the r.h.s. of this equation is zero. Then, the $\vartheta$ field generates the additional contribution to the energy-momentum tensor $T^{\mu\nu}$, and hence, to the r.h.s. of (\ref{mee}).

Now, we present, first, some classical solutions for the CS modified gravity, second, the methodology allowing the gravitational CS term as a quantum correction.

\subsection{Classical solutions}

So, our task will consist in solving the equations (\ref{mee}) with the additional condition (\ref{conscond}). As a first example, we consider a static spherically symmetric metric \cite{JaPi}:
\begin{equation}
\label{SSSM}
ds^2=N^2(r)dt^2-A^2(r)dr^2-r^2d\Omega^2.
\end{equation}
This is a very broad class of metrics including Schwarzschild, Reissner-Nordstr\"{o}m and many other metrics. As we already said, in this case the non-zero components of the curvature tensor are $R_{[ab][ab]}$, so, the consistency condition (\ref{conscond}) is automatically satisfied. For this metric, one has only non-zero components of the Ricci tensor $R^r_r=\frac{A^{\prime}}{rA^2}$, $R^{\theta}_{\theta}=\frac{1}{r^2}(1-\frac{1}{A})+\frac{A^{\prime}}{rA^2}$.
Then, we can consider the vacuum case $T^{\mu\nu}=0$, and choose the vector $v_{\mu}=\partial_{\mu}\vartheta$ to be purely timelike, $v_{\mu}=(\frac{1}{\mu},\vec{0})$, with $\mu=const$, i.e. $\vartheta=\frac{t}{\mu}$. In this case, the components $C^{00}$ and $C^{0i}=C^{i0}$ of the Cotton tensor immediately vanish \cite{JaPi}. A bit more involved calculation (see details in \cite{JaPi}) allows to show that the $C^{ij}$ components also vanish. As a result, we conclude that the spherically symmetric static solutions of the usual Einstein equations solve the modified equations (\ref{mee}) as well. It is clear that if one suggests the $\vartheta$ to be dynamical, the equation (\ref{eqtheta}) for $\vartheta$ will be satisfied if the potential is zero, and $\vartheta=\frac{t}{\mu}$. We note that this choice for $\vartheta$ is a particular case of the expression $\vartheta=k_{\mu}x^{\mu}$ used within studies of the Lorentz symmetry breaking in CSMG which we will discuss further. 

Moreover, it has been shown in \cite{Grumiller} that all, even non-static ones, spherically symmetric metrics given by
\bea
\label{SSM}
ds^2=g_{\mu\nu}(x^{\lambda})dx^{\mu}dx^{\nu}+\Phi^2(x^{\rho})d\Omega^2,
\eea
where $d\Omega^2$ is the 2-sphere line element, so that the coordinates on the sphere are $x^i$, and $x^{\mu}$ are two remaining coordinates (one of them is necessarily timelike), solve the modified Einstein equations (\ref{mee}) for 
\bea
\vartheta=F(x^{\mu})+\Phi(x^{\mu})G(x^i),
\eea
where $G(x^i)$ and $F(x^{\gamma})$ are the arbitrary functions of sphere coordinates and remaining coordinates respectively, and $\Phi$ is defined in (\ref{SSM}). The class of spherically symmetric metrics (\ref{SSM}) involves not only the static ones (\ref{SSSM}) but also many other metrics, including the FRW cosmological metric (the cosmological aspects of CSMG were also discussed in many papers, f.e. in \cite{Soda}). Some types of metrics with cylindrical symmetry were also shown in \cite{Grumiller} to be consistent within the CSMG.

Now, let us discuss the consistency of the G\"{o}del-type metric (\ref{gtype}) in CSMG. We consider the equations of motion (\ref{mee}) in the tetrad base, following \cite{ourCS}.

In the non-dynamical case, with appropriate choice of units, the equations (\ref{mee}) imply
\begin{eqnarray}
R_{AB}&+&C_{AB}=\kappa(T_{AB}-\frac{1}{2}\eta_{AB}T)+\Lambda\eta_{AB};\\
C^{AB}&=&-\frac{1}{2}[\varepsilon^{CADE}(\nabla_DR^B_E)\partial_C\vartheta+{}^*R^{EAFB}\nabla_E\nabla_F\vartheta]+(A\leftrightarrow B).\nonumber
\end{eqnarray}
The divergence of modified Einstein equations is
\begin{equation}
\nabla_AC^{AB}=\frac{1}{8}{}^*RR\partial^B\vartheta.
\end{equation}
In tetrad base, the components of Ricci tensor for G\"{o}del-type metric are constant, which is an essential advantage of this base. Actually, one has
\begin{equation}
R_{00}=2\omega^2,\quad\, R_{11}=R_{22}=2\omega^2-m^2,\quad\, R=2(\omega^2-m^2).
\end{equation}
Following the methodology described in \cite{RebTio}, we consider three cases of $H$ and $D$ consistent with the conditions of space-time homogeneity of the metric (\ref{condhom}):\\
(i) hyperbolic, $H=\frac{2\omega}{m^2}[\cosh mr -1]$, $D=\frac{1}{m}\sinh mr$;\\
(ii) trigonometric, $H=\frac{2\omega}{\mu^2}[1-\cos \mu r]$, $D=\frac{1}{m}\sin \mu r$; $\mu^2=-m^2$;\\
(iii) linear, $H=\omega r^2$, $D=r$.\\
Repeating the argumentation from \cite{RebTio}, one immediately sees that for $0<m^2<4\omega^2$, there is a noncausal region with $r>r_c$, where $\sinh^2\frac{mr_c}{2}=(\frac{4\omega^2}{m^2}-1)$.\\
So, at $m^2\geq 4\omega^2$ there is no problems with causality.

Now let us choose the matter. We have three most important its examples \cite{ourCS,RebTio}:

\noindent (i) Fluid, $T_{AB}=(\rho+p)u_Au_B+p\eta_{AB}$, $u^A=(1,0,0,0)$, $T_{00}=\rho$, $T_{11,22,33}=p$.\\
(ii) Scalar, $\psi=s(z-z_0)$, $T_{00,33}=\frac{s^2}{2}$, $T_{11,22}=-\frac{s^2}{2}$.\\
(iii) Electromagnetism, $F_{03}=-F_{30}=e\sin[2\Omega(z-z_0)]$, $F_{12}=-F_{21}=-E\cos(2\Omega(z-z_0))$,
$T_{00,11,22}=\frac{e^2}{2}$, $T_{33}=-\frac{e^2}{2}$.

The matter can be presented by composition of these three types. Then, the non-zero components of the Cotton tensor in this base look like
\begin{eqnarray}
C_{00}&=&2\frac{\partial\vartheta}{\partial z}\omega(4\omega^2-m^2);\quad\, C_{11}=C_{22}=\frac{1}{2}C_{00};\nonumber\\
C_{01}&=&-\frac{1}{2}\frac{\partial^2\vartheta}{\partial z\partial t}\frac{H}{D}(4\omega^2-m^2);\nonumber\\
C_{02}&=&-\frac{1}{2}\frac{\partial^2\vartheta}{\partial z\partial r}(4\omega^2-m^2);\nonumber\\
C_{03}&=&-\frac{1}{2}\frac{\partial\vartheta}{\partial t}\omega(4\omega^2-m^2);\nonumber\\
C_{13}&=&-\frac{1}{2}\frac{\partial^2\vartheta}{\partial t^2}\frac{H}{D}(4\omega^2-m^2);\nonumber\\
C_{23}&=&\frac{1}{2}\frac{\partial^2\vartheta}{\partial r\partial t}(4\omega^2-m^2).
\end{eqnarray}
It is clear that the Cotton tensor is traceless, $C_A^A=0$.
To cancel the off-diagonal components of $C_{AB}$ we choose $\vartheta(z)=b(z-z_0)$ which matches the suggestion done above that the vector $v_M=\partial_M\vartheta$ is constant, which will be further used to study the Lorentz symmetry breaking. 
We introduce also $k=b\omega$, and require $4\omega^2\neq m^2$.

The system of the modified Einstein equations (for 00, 11=22, 33 components respectively) looks like:
\begin{eqnarray}
2\,\omega^{2}+2\,b\omega(4\,\omega^{2}-m^{2}) & = & \frac{1}{2}\,e^{2}+\frac{1}{2}\,\rho-\Lambda+\frac{3}{2}\,p,\label{eq:cs00}\\
2\,\omega^{2}-m^{2}+b\omega(4\,\omega^{2}-m^{2}) & = & \frac{1}{2}\,e^{2}-\frac{1}{2}\,p+\Lambda+\frac{1}{2}\,\rho,\nonumber\\
0 & = & -\frac{1}{2}\,e^{2}-\frac{1}{2}\,p+s^{2}+\Lambda+\frac{1}{2}\,\rho.\nonumber
\end{eqnarray}
We note, that, just as in the Einstein case \cite{RebTio}, this system is a purely algebraic one.
Let us solve these equations. After some manipulations we arrive at equations for $m^2$ and $\omega^2$, with $k=b\omega$ (we note that at $b=0$, the usual GR solution is replayed since in this case, $\vartheta=0$!):
\begin{eqnarray}
(2+8\,k)\omega^{2}-2\,km^{2}  &= & \rho+s^{2}+p,\\
\left(2+4\,k\right)\omega^{2}-\left(1+k\right)m^{2} & = & -s^{2}+e^{2}.
\end{eqnarray}
One of the interesting new results having no GR analogue is the vacuum noncausal solution $m^2=\omega^2$, $b=-\frac{1}{3\omega}$, $\Lambda=0$. Some other interesting conclusions of the above system are that, unlike the general relativity, the hyperbolic causal solutions are possible in CS modified gravity, and that trigonometric and linear solutions can arise only for a non-zero electromagnetic field \cite{ourCS}.

If one suggests that the CS coefficient is dynamical, more new solutions having analogues neither in GR nor for the case of the non-dynamical CS coefficient are possible, see details in \cite{ourCS}, with again the Einstein equations will be reduced to the algebraic equations involving some extra additive terms in comparison with (\ref{eq:cs00}). In particular, one can have a vacuum solution, where only cosmological constant is non-zero while density, pressure and all fields are zero.

At  the same time, it is necessary to emphasize that not any solution consistent in the GR will be consistent also in CS modified gravity. The paradigmatic example is the Kerr metric which fails to solve new equations of motion \cite{JaPi,AlYunes}. It has been shown then in \cite{YunesPre} that, to satisfy the modified Einstein equations in the dynamical CS modified gravity, the Kerr metric should be also modified, by adding the $\vartheta$-dependent terms, with the equations of motion are afterwards solved order by order in $\vartheta$. Clearly, studies of consistency of various metrics possessing no rotational symmetry within the CS modified gravity represent an open problem.

To close the discussion of the classical solutions, it is necessary to discuss the propagation of the plane waves.
Similarly to the Section 2.2, we introduce the transverse-traceless components $h^{TT}_{ij}$ which are the only physical variables in the theory (so, there are only two independent components, that is, if the plane wave propagates f.e. along $x_3$, we have only $h_{11}=-h_{22}=T$ and $h_{12}=h_{21}=S$). 

In this case, for the time-like vector $v_{\mu}=(\mu^{-1},0,0,0)$ the quadratic Lagrangian takes the form:
\begin{equation}
L_2=-\frac{1}{4}h_{ij}^{TT}\Box h_{ij}^{TT}+\frac{1}{4\mu}\epsilon^{ijk}h_{il}^{TT}\Box\partial_kh_j^l+\ldots,
\end{equation}
where dots are for physically irrelevant (non-propagating) degrees of freedom.

The corresponding linear equation of motion is
\begin{equation}
-\frac{1}{2}\Box h^{ij}_{TT}+\frac{1}{2\mu}\epsilon^{ilk}\Box \partial_k h^j_{l,TT}=0.
\end{equation}
As a result, one immediately concludes that the dispersion relation is the usual one, $k^2_0=\vec{k}^2$, and both polarizations propagate with the speed of light.

The natural question is -- what is difference of these polarizations? A more careful analysis \cite{JaPi} shows that, for plane waves proportional to $e^{i\omega t-ikz}$, one finds that there are two basic (circular) polarizations $T=iS$ and $T=-iS$, with their intensities proportional to $(1+\frac{k}{\mu})^{-2}$ and $(1-\frac{k}{\mu})^{-2}$ respectively. This difference of intensities can be treated as a consequence of parity breaking.

It should be noted that if we consider, instead of the CS term, the one-derivative additive term
$h_{\mu\nu}\epsilon^{\lambda\alpha\mu\rho}\theta_{\lambda}\partial_{\rho}h^{\nu}_{\alpha}$, with $\theta^{\lambda}$ being a space-like vector, we will have  two polarizations with physically consistent dispersion relations $E=\pm \theta+\sqrt{p^2+\theta^2}$, so, in this case the velocities differ from speed of light \cite{ourlingrav}. However, this term is not gauge invariant, which, within the gravity context, means that it breaks the general covariance.

\subsection{Perturbative generation}  

The special interest is attracted to the gravitational CS term within the context of study of the Lorentz symmetry breaking. The main reason consists in the fact that, besides of the CPT symmetry breaking, for a special choice of the CS coefficient $\vartheta=b_{\mu}x^{\mu}$, where $b_{\mu}$ is a constant vector (as we already noted in the previous subsection, this choice is consistent with the G\"{o}del-type solutions), the CS term displays Lorentz symmetry breaking, taking the form (\ref{fullCS03}), or, for the weak field, the linearized form (\ref{fullCS04}). Therefore the natural idea consists in a generation of this term as a perturbative correction, similarly to the generation of the CFJ term in the extended QED, see f.e. \cite{JK}. This similarity is supported by a natural analogy between the gravitational anomalies \cite{AlvAnom} and the Adler-Bell-Jackiw (ABJ) anomaly \cite{ABJ}. Moreover, it follows from \cite{JackAmb} that this anomaly is deeply related with the ambiguity of results, therefore, it is natural to expect the ambiguity of the gravitational CS term as well.

So, one can start with the action of spinors coupled to gravity, where the Lorentz-breaking vector $b_{\mu}$ is introduced:
\begin{equation}
\label{laggra}
S=\int d^4x e\bar{\psi}(i\ds-m-\bs\gamma_5+\os)\psi,
\end{equation}
here, $\bs=b^{\mu}e_{\mu}^a\gamma_a$, and $\omega_{\mu}=\frac{1}{4}\omega_{\mu bc}\sigma^{bc}$ is a (Riemannian) connection. We note that the CS term dominates in the limit $m\to 0$ while the one-derivative term discussed in \cite{ourlingrav} vanishes in this limit.
The corresponding one-loop effective action is given by the following trace of the logarithm:
\begin{equation}
\label{trace}
\Gamma^{(1)}=i{\rm Tr}\ln(i\ds-m-\bs\gamma_5+\os).
\end{equation}
Just the same approach was used in \cite{JK} for the Lorentz-breaking extension of QED.
In the weak gravity case, we can use the approximation $e_{\mu a}\simeq\eta_{\mu a}+\frac{1}{2}h_{\mu a}$. This trace of the logarithm, however, can be calculated both in the weak field case and in the full-fledged gravity case, with use of the Feynman diagrams or of the proper-time method.

It is interesting that, similarly to the CFJ term, the $4D$ gravitational CS term is ambiguous, i.e. the results for it depend on the calculation scheme. So, within all these approaches, the linearized gravitational CS term 
\begin{equation}
\label{tzero}
S_{CS}= C \int d^4x h_{\mu\nu} \epsilon^{\mu \rho \kappa \lambda} b_\kappa \partial_\lambda \left(\Box h_{\rho}\,\!^{\nu}-\partial^{\nu} \partial^{\sigma} h_{\rho\sigma}\right),
\end{equation}
or its full-fledged analogue (\ref{fullCS03}) multiplied by $2C$, was shown to arise, with the constant $C$ depends on the method of computation. So, in \cite{CSfirst}, where the calculations were carried out in the weak gravity case with use of the Feynman diagrams constructed for the action (\ref{laggra}), it was found that $C=\frac{1}{192\pi^2}$. Further, in \cite{finT}, this scheme has been realized for the finite temperature case where the zero component of the internal momentum is supposed to be discrete, $k_0=(2n+1)\pi T$, so that the result is
\begin{eqnarray}
\label{resultT}
S_{CS}&=&  \int d^4x\, h_{\mu\nu} \left[\frac{1}{192 \pi ^2}\epsilon^{\rho\mu \kappa \lambda} b_\kappa \partial_\lambda \left(\Box h_{\rho}\,\!^{\nu}-\partial^{\nu} \partial^{\sigma} h_{\rho\sigma}\right)\right. \\
&+& \left. \frac{T^2}{12} b_0 \epsilon^{\rho\mu \kappa \lambda} u_\kappa \partial_\lambda \left(\frac{\partial_0 \partial^{\nu}}{\Box} -u^{\nu}\right)\left(\frac{\partial_0 \partial^{\sigma}}{\Box} -u^{\sigma} \right)h_{\rho\sigma}\right],\nonumber
\end{eqnarray}
i.e. it looks like a sum of the zero-temperature result (\ref{tzero}) and the additive term proportional to $T^2$.

In \cite{ptime}, where the proper time method has been used for the full-fledged gravity, the result was found in the form (\ref{fullCS03}), with $C=\frac{1}{128\pi^2}$. Finally, in \cite{ourFuji} it has been argued that due to the arbitrariness in defining of conserved currents within the functional integral approach, the constant $C$ is actually completely ambiguous. The similar situation occurs in QED \cite{Chung}. However, the ambiguity of results is known to be highly controversial, and in gravity it is even more controversial than in electrodynamics. For example, in \cite{Alts} it was claimed that, if one suggests that the $b_{\mu}$ is the vacuum expectation value (v.e.v.) of a some dynamical field, the correct result for the $4D$ gravitational CS term is zero, as is also required by the gauge invariance of the Lagrangian (and not only the action). Nevertheless, the question whether the requirements of \cite{Alts} are indeed so necessary is still open, as the presence of ambiguities in generic Lorentz-breaking theories is a strongly polemical problem.

However, there are also other interesting scalar-tensor gravity models which we will consider now.

\section{Brans-Dicke gravity}

The Brans-Dicke (BD) gravity is one of the most known and studied scalar-tensor gravity models.  Originally, it has been introduced in \cite{BD}, basing on the idea that the physical space itself possesses geometrical features beyond those ones generated by matter (this is one of the forms of the so-called Mach principle), so, the action of the BD gravity was proposed in the form
\begin{equation}
\label{BD}
S=\int d^4x\sqrt{|g|}\Big(\phi R+\frac{\omega}{\phi}\partial_a\phi\partial^a\phi+16\pi {\cal L}_{mat}).
\end{equation}
In this theory, the new scalar field $\phi$ (which does not contribute to the matter Lagrangian) plays the role of the effective gravitational constant; indeed, if one chooses $\phi=\frac{1}{2\kappa^2}$, the theory reduces to the Einstein gravity with the usual matter. One advantage of the theory consists in the fact that the coupling constant $\omega$ is dimensionless, hence the negative-dimension constants jeopardizing the renormalizability of the gravity are ruled out. Also, in this case the gravitational constant has a dynamic origin being related with an asymptotic value of the $\phi$.

For this theory, one can derive equations of motion:
\bea\label{3.2}
&-&\frac{2\omega}{\phi}\Box\phi+\frac{\omega}{\phi^{2}}\partial_{\mu}\phi\partial^{\mu}\phi+R=0;\\
\label{3.3}
&& R_{\mu\nu}-\frac{1}{2}g_{\mu\nu}R=\left(\frac{8\pi}{\phi}\right)T_{\mu\nu}-\frac{\omega}{\phi^{2}}\left(\partial_{\mu}\phi\partial_{\nu}\phi
-\frac{1}{2}g_{\mu\nu}\partial_{\rho}\phi\partial^{\rho}\phi\right)
+
\frac{1}{\phi}\left[\nabla_{\nu}(\partial_{\mu}\phi)-g_{\mu\nu}\Box\phi\right],\nonumber
\eea
where $T_{\mu\nu}$ is the energy-momentum tensor of the usual matter (not including $\phi$).
Contracting this equation with $g^{\mu\nu}$, we find
\begin{equation}\label{3.31} 
R=-\left(\frac{8\pi}{\phi}\right)T-\frac{\omega}{\phi^{2}}\partial_{\rho}\phi\partial^{\rho}\phi+\frac{3}{\phi}\Box\phi,
\end{equation}
which we can combine with the Eq. (\ref{3.2}), obtaining
\begin{equation}\label{3.32} 
\Box\phi=\left(\frac{8\pi}{3-2\omega}\right)T.
\end{equation}
The equations (\ref{3.31},\ref{3.32}) are analogues of the Einstein equations and can be solved.
 
 As a first example, we consider the static spherically symmetric metric (\ref{SSSM}) which we now rewrite as
 \begin{equation}
ds^2=e^{2\alpha(r)}dt^2-e^{2\beta(r)}(dr^2+r^2d\Omega^2)
\end{equation}
In the vacuum case, $T_{\mu\nu}=0$, this metric will be a consistent solution of equations of motion \cite{BD}. Explicitly, one finds
\begin{eqnarray}
e^{\alpha(r)}&=&e^{\alpha_0}\left[\frac{1-\frac{2B}{r}}{1+\frac{2B}{r}}
\right]^{1/\lambda};\nonumber\\
e^{\beta(r)}&=&e^{\beta_0}(1+\frac{2B}{r})^2\left[\frac{1-\frac{2B}{r}}{1+\frac{2B}{r}}
\right]^{(\lambda-C-1)/\lambda};\\
\phi(r)&=&\phi_0e^{\alpha_0C}\left[\frac{1-\frac{2B}{r}}{1+\frac{2B}{r}}
\right]^{C/\lambda}.\nonumber
\end{eqnarray}
The cosmological solutions also were found in \cite{BD} where they were shown, in the vacuum case, to look like
\bea
\phi&=&\phi_0 t^r, \quad\, a=a_0t^q;\nonumber\\
r&=&\frac{2}{4-3\omega},\quad\, q=\frac{2-2\omega}{4-3\omega},
\eea
so, accelerating solutions ($q>1$) are possible for $\omega>2$. Further, various papers, continuing this study, discussed cosmic acceleration in BD gravity in details, see f.e. \cite{Sen}.

Now, let us discuss the G\"{o}del-type solutions (\ref{gtype}) in the BD gravity. It has been shown in \cite{ourBD} that the nontrivial solution, i.e. that one with a non-constant scalar $\phi$ (otherwise the BD gravity reduces trivially to the Einstein gravity) is possible only if the action (\ref{BD}) includes the cosmological constant as well, so, one has
\begin{equation}
\label{BD1}
S=\int d^4x\sqrt{|g|}\Big(\phi (R-2\Lambda)+\frac{\omega}{\phi}\partial_a\phi\partial^a\phi+16\pi {\cal L}_{mat}).
\end{equation}
The modified Einstein equations, in the tetrad base, look like
\begin{eqnarray}\label{4.19}
G{^{A}_{B}}-\delta{^{A}_{B}}\Lambda&=&
\left(\frac{8\pi}{\phi}\right)T{^{A}_{B}}-\frac{\omega}{\phi^2}\left(\partial^{A}\phi\partial_{B}\phi-\frac{1}{2}\delta{^{A}_{B}}\partial_{C}\phi\partial^{C}\phi
\right)+\nonumber\\ &+&\phi^{-1}\left(\nabla_{B}\partial^{A}\phi
-\delta{^{A}_{B}}\Box\phi\right),
\end{eqnarray}
and choosing again the matter in the form of a composition of the fluid and electromagnetic field (see Section 3.2.2), with the angular velocity parametrizing the G\"{o}del-type metric (\ref{gtype}) and defined within the conditions (\ref{condhom}) is now denoted as $\Omega$ instead of $\omega$, we find that the case $\phi=\phi(z)$ yields
\begin{equation}\label{5.38}
4\Omega^{2}-m^{2}=\left(\frac{8\pi}{\phi}\right)(\rho+E_{0}^2), \quad m^2+2\Lambda=-\frac{\phi''}{\phi}.
\end{equation}
The typical cases are:\\
(i) $4\Omega^2-m^2=0$ (causal solution!), $\rho+E^2_0=0$. In this case $\phi$ is a trigonometric function.\\
(ii) $\rho=const$, $\phi=const$ -- trivial case reducing to GR.

For $\phi=\phi(t)$, one arrives at $\phi=const$, and this case is also trivial. In principle, more involved situations can be studied as well. As for the black hole solutions in BD gravity, we strongly recommend the classical paper \cite{HawkBD}. In principle, many other solutions for the BD gravity have been studied, including global monopoles, wormholes etc., but the limited volume of these notes does not allow for their detailed discussion.
 
\section{Galileons} 

One of the most important examples of the scalar-tensor gravity models is the galileons theory proposed originally in \cite{Hornd}. Its key idea is as follows: let us consider the most general scalar-tensor action involves no more than second derivatives of the metric tensor and no more than the first ones of the scalar field. Effectively, it was a suggestion of the Lovelock-like construction not only in the gravitational sector but also in the scalar one. So, we suggest the action to look like
\bea
\label{gal}
S=\int d^4x\sqrt{-g}{\cal L}(g_{\mu\nu}, \pa_{\lambda} g_{\mu\nu}, \pa_{\lambda}\pa_{\rho}g_{\mu\nu};\phi,\pa_{\mu}\phi).
\eea
As a result, the equations of motion involve various tensors constructed on the base of the Riemann curvature and its covariant derivatives, and various derivatives of the scalar field. In principle we can have the gravity equations of motion with Lovelock-like l.h.s. and non-canonical scalar-dependent r.h.s., and strongly nonlinear equations of motion for the scalar. We note that there is no ghost problem here since there is no higher derivatives. In principle, even on the flat background, one can have a theory of a scalar field with highly nonlinear equation of motion, the so-called $K$-theory (see \cite{ktheory} and references therein). 

However, the model (\ref{gal}) was forgotten for a long time and revitalized only in 2008, in the paper \cite{Ratt} where the concept of galileons was formulated.  Its key idea consists in invariance of the theory with respect to the combination of dilatations and conformal transformations so that the new scalar $\pi$ varies as $\pi\to\pi+c+b_{\mu}x^{\mu}$, where $c$ and $b_{\mu}$ are constants. These transformations look similarly to the Galilean ones, therefore the $\pi$ was called the galileon. So, again, the key idea is that we have derivative couplings but no higher derivatives in the kinetic term.

There are five terms with the symmetry above. Let us introduce notations $\Pi^{\mu\nu}=\pa^{\mu}\pa^{\nu}\pi$,  $[A]=A_{\mu}^{\mu}$ for trace (so, $\frac{1}{2}[\Pi]\pa\pi\cdot\pa\pi=\frac{1}{2}\Box\pi\pa^{\mu}\pi\pa_{\mu}\pi$), $[\Pi]=\Box\pi$, etc.), and use a dot for the usual scalar product like $A\cdot B\equiv A_{\mu}B^{\mu}$.
So, we can write our five terms as:
\bea
{\cal L}_1&=&\pi,\nonumber\\
{\cal L}_2&=&-\frac{1}{2}\pa\pi\cdot\pa\pi;\nonumber\\
{\cal L}_3&=&-\frac{1}{2}[\Pi]\pa\pi\cdot\pa\pi;\nonumber\\
{\cal L}_4&=&-\frac{1}{4}\Big([\Pi]^2 \pa\pi\cdot\pa\pi-2[\Pi]\pa\pi\cdot\Pi\cdot\pa\pi-[\Pi^2]\pa\pi\cdot\pa\pi+
2\pa\pi\cdot\Pi^2\cdot\pa\pi\Big);\nonumber\\
{\cal L}_5&=&-\frac{1}{5}\Big([\Pi]^3\pa\pi\cdot\pa\pi-3[\Pi]^2\pa\pi\cdot\Pi\cdot\pa\pi-3[\Pi][\Pi^2]\pa\pi\cdot\pa\pi+\nonumber\\ 
&+& 6[\Pi]\pa\pi\cdot\Pi^2\cdot\pa\pi+2[\Pi]^3\pa\pi\cdot\pa\pi+3[\Pi^2]\pa\pi\cdot\Pi\cdot\pa\pi-
6\pa\pi\cdot\Pi^3\cdot\pa\pi
\Big).
\eea
The complete Lagrangian of $\pi$ is a linear combination of these terms: ${\cal L}=c_1{\cal L}_1+c_2{\cal L}_2+c_3{\cal L}_3+c_4{\cal L}_4+c_5{\cal L}_5$. Clearly, the next step consists in coupling of these Lagrangians to gravity. But let us first describe some perturbative effects of these couplings.

One of the interesting effects is that these galileon terms ${\cal L}_i$ are not renormalized under quantum corrections! The reasons are as follows \cite{TrodGal}. First, the galileon is massless, so, its propagator is $1/k^2$. Then, all galileon couplings $c_3,c_4,c_5$ have negative mass dimensions, therefore the contributions to these terms possess quadratic and even higher divergences.  After integration of subloops, the leading divergence is proportional to $\int d^4k(k^2)^n$, with $n\geq -1$, and this integral vanishes within dimensional regularization. Finally, the subleading contributions to galileon vertices vanish as well (this proof is more sophisticated being based on analysis of symmetries). In principle, such conclusions are natural for a massless theory with derivative couplings. Other divergent contributions in the galileons theory in the flat space, which do not match the form of the classical action, in particular, involve more derivatives (f.e. $\Box^2$ terms), are discussed in \cite{Lavi}.

Clearly, the next step is the coupling of the scalar $\pi$ to the gravity. One of the first ideas consists in coupling of galileons to the curvature, so we have terms like \cite{Deffa,Appleby}:
\bea
\delta S_4=\int d^4x \sqrt{-g}(\pi_\mu\pi^{\mu})(\pi_{\nu}G^{\nu\rho}\pi_{\rho}),
\eea
where $\pi_{\mu}\equiv\nabla_{\mu}\pi$, etc., or the higher terms like $\pi_{\mu}\pi^{\mu\nu}\pi^{\rho}G_{\nu\rho}$, or the simplest terms $\pi^{\mu}\pi^{\nu}G_{\mu\nu}$ (the last term is the example of the John term, see below).  So, effectively we have a gravity-coupled scalar field with strongly nonlinear dynamics involving derivative depending couplings. As it has been claimed in \cite{Appleby}, these terms are of special interest within the cosmological context, where it has been explicitly shown that the solutions with constant $H=\frac{\dot{a}}{a}$ are consistent for the presence of galileons, therefore de Sitter-like exponential expansion is possible in this case, with neither potential term for the scalar nor cosmological constant are employed, therefore the galileons theory is a sound candidate for the role of the dark energy. In \cite{Charm}, it was argued that only the minimal scalar-gravity couplings must be considered, as a result, there were introduced four typical galileon-gravity coupling terms called John, Paul, George and Ringo:
\bea
{\cal L}_{John}&=&V_J(\pi)G_{\mu\nu}\nabla^{\mu}\pi\nabla^{\nu}\pi;\nonumber\\
{\cal L}_{Paul}&=&V_P(\pi)P_{\mu\nu\rho\sigma}\nabla^{\mu}\pi\nabla^{\nu}\pi\nabla^{\rho}\nabla^{\sigma}\pi;\nonumber\\
{\cal L}_{George}&=&V_G(\pi)R;\nonumber\\
{\cal L}_{Ringo}&=&V_R(\pi){\cal G}.
\eea
 where $P^{\mu\nu\alpha\beta}=-\frac{1}{4}\epsilon^{\mu\nu\rho\sigma}\epsilon^{\alpha\beta\gamma\delta}R_{\rho\sigma\gamma\delta}$ is the double dual of the Riemann curvature.  In \cite{Charm}, the cosmological aspects of the theory involving these terms were studied, especially, it was argued how the known cosmological self-tuning problem is solved in this theory. Various issues related to the cosmic acceleration in this context are studied numerically also in \cite{StaroGal}. Many other papers are also devoted to galileon cosmology. However, up to now the galileons are mostly considered namely within the cosmological context, there are only a few papers on other solutions such as f.e. black holes (see f.e. \cite{Rinaldi}). An interesting review of galileons is presented in \cite{Curtright}. To close this section, we note that many aspects of galileons still must be studied.
   
\section{Conclusions}   

We formulated several examples of scalar-tensor gravity models whose form does not match the standard quintessence-gravity Lagrangian ${\cal L}=\sqrt{|g|}(\frac{1}{16\pi G}R-\frac{1}{2}g^{\mu\nu}\pa_{\mu}\phi\pa_{\nu}\phi-V(\phi))$ which is well studied, both within the cosmological and QFT contexts. Explicitly, we considered the $4D$ CS modified gravity, the Brans-Dicke gravity and the galileons theory.  These theories display new interesting features.

 First of all, the CSMG allows for the CPT symmetry breaking, and, for a certain form of the CS coefficient, also for the Lorentz symmetry breaking, opening thus a way for intensive studies of Lorentz-breaking modifications of gravity. Some of these studies will be discussed in the next chapter. Besides, in the presence of the gravitational CS term new solutions impossible within the usual GR arise. 

Second, the Brans-Dicke gravity represents itself as a theory allowing to rule out the gravitational constant possessing negative mass dimension and hence implying in problems with quantum description of the gravity. Moreover, it turns to be that some new solutions which are not consistent within the GR, are also possible. 

Third, the galileons theory turns out to be a sound candidate for a description of the dark energy allowing for accelerated solutions. Besides of this, the galileons contributions to the action arise within applying the Stuckelberg approach for the massive gravity. Essentially, at the first step one introduces the new vector field to construct the gauge invariant extension for the mass term of the gravity, and at the second step, to achieve the gauge symmetry for this vector field, one introduces the scalar field whose action matches the galileon form \cite{Hinter}. 

To conclude, for the scalar-tensor gravity models, one has essentially new results. One of the most interesting conclusions is the possibility to introduce the Lorentz symmetry breaking within the gravitational context, for a special form of the CS coefficient. However, it is clear that in this context, an extension of gravity through introduction of vector fields seems to be more promising since the vacuum expectations of vector fields can yield constant vectors necessary to introduce privileged space-time directions breaking thus the Lorentz symmetry. 
   
\chapter[Vector-tensor gravities]{Vector-tensor gravities and problem of Lorentz symmetry breaking in gravity}

\section{Introduction and motivations}

The interest to vector-tensor gravity models strongly increased in recent years. One of the main motivations to studying these models arises from the idea of the Lorentz symmetry breaking.  Indeed, as it is well known, in the flat space the explicit Lorentz symmetry breaking is implemented through introduction of a constant vector (tensor) generating a space-time anisotropy (see f.e. \cite{ColKost,KosGra}). As we already noted in the previous chapter, this methodology allowed to define, for example, the Carroll-Field-Jackiw term (\ref{CFJ}) as well as many other terms discussed in \cite{ColKost}. However, in the curved space the explicit Lorentz symmetry breaking faces serious problems. First of all, the definition of the constant vector (tensor) itself in this case becomes highly controversial: for example, while in the flat space the constant vector $k^{\mu}$ is defined to satisfy the condition $\partial_{\nu}k^{\mu}=0$, this condition cannot be applied in a curved space since it breaks the general covariance. The possible "covariant extension" of this condition like $\nabla_{\nu}k^{\mu}=0$ would imply in extra restrictions for the space-time geometry (and, moreover, nobody could guarantee these restrictions to be satisfied for a general choice of the vector $k_{\mu}$). In principle, one can also deal with derivative expansions of corresponding effective actions, where various orders of derivatives of "constant" tensors can be obtained (see f.e. \cite{Berr}), however, it is clear that in this case the definition of a constant vector (or tensor) simply loses its sense, and such a vector becomes an extra field. Moreover, in many cases such possible new terms are not gauge invariant which means that together with the Lorentz symmetry, the general covariance for such terms is broken as well (the problem of breaking the general covariance in modified gravity is discussed in details in \cite{KosMew}; in principle, it should be noted that breaking of general covariance occurs for the term $u^{\mu}u^{\nu}R_{\mu\nu}$ proposed in \cite{CarTam} as a possible example of a CPT-even Lorentz-breaking term for gravity, as well as for the one-derivative linearized term discussed in \cite{ourlingrav}).

Therefore, the most appropriate method for implementing the Lorentz symmetry breaking into a curved space-time turns out to be based on the spontaneous symmetry breaking. Its essence is as follows. One considers the action of the metric tensor coupled to the vector field (again, similarly to the previous chapter, this vector field is treated as an ingredient of gravity model itself but not as the matter, thus, we have the vector-tensor gravity) so that the purely metric sector is presented by the usual Einstein-Hilbert action, and the dynamics of the vector field is described by the Maxwell-like term, plus a potential whose minimum yields a vector implementing the Lorentz symmetry breaking, and maybe also some extra terms responsible for a vector-gravity coupling. The paradigmatic example is the bumblebee action \cite{KosBluhm} (the name "bumblebee" itself was introduced in \cite{Bertolami}), looking like
\bea
\label{bumblebee}
S=\int d^4x\sqrt{|g|}\left(\frac{1}{16\pi G}(R+\xi B^{\mu}B^{\nu}R_{\mu\nu})-\frac{1}{4}B_{\mu\nu}B^{\mu\nu}-V(B^{\mu}B_{\mu}\pm b^2)
\right).
\eea
Here $\xi$ is a dimensionless constant, $B_{\mu\nu}=\partial_{\mu}B_{\nu}-\partial_{\nu}B_{\mu}$ is the stress tensor for the bumblebee field $B_{\mu}$, and $V$ is the potential possessing an infinite set of minima $B_{0\mu}$ satisfying the condition $B^{\mu}_0B_{0\mu}=\pm b^2$ (the difference of signs reflects that the vector $B_{0\mu}$ can be either time-like or space-like, while $b^2>0$). So, actually choosing of one of the vacua $B_{0\mu}$ allows to introduce the privileged direction. The potential is usually chosen to be quartic in the field $B_{\mu}$ by renormalizability reasons. Alternatively, one can deal with Einstein-aether theory where, instead of this, the minima arise due to a constraint multiplied by a Lagrange multiplier $\sigma$, so that one has $V=\sigma(B^{\mu}B_{\mu}\pm b^2)$, but the kinetic term is not Maxwell-like being a more generic quadratic function of covariant derivatives of the vector $B_{\mu}$. In principle, one can consider the vector-tensor gravity models without any potential \cite{Beltran}, however, in this case the spontaneous Lorentz symmetry breaking cannot occur. Such theories are considered mostly within the cosmological context (see f.e. \cite{Beltran}).

Within this chapter, we discuss some interesting classical results for the Einstein-aether gravity and for the bumblebee gravity. At the end of the chapter, we also will review some terms proposed in \cite{ColKost,KosGra} as possible extensions of  the Einstein gravity allowing to break the Lorentz symmetry explicitly. As for the Horava-Lifshitz gravity, although it represents itself as an example of non-Lorentz-invariant gravity model, it is described in terms of the essentially distinct methodology and will be discussed in the next chapter.

\section{Einstein-aether gravity}
 
 So, let us implement the spontaneous Lorentz symmetry breaking in a curved space-time. To justify importance of this approach, one can remind that namely the spontaneous breaking mechanism has been initially proposed to explain the origin of the Lorentz symmetry breaking in the low-energy limit of the string theory \cite{KosSam}. Following this concept, one considers a vector field $B_{\mu}$ with a constant square, i.e. $B^{\mu}B_{\mu}=\pm b^2$, which is implemented via introducing the constraint with use of the Lagrange multiplier $\sigma$, adding to the Lagrangian the potential $V=\sigma(B^{\mu}B_{\mu}\pm b^2)$. Alternatively, as we already noted above, one can introduce the quartic potential. The approach based on the Lagrange multiplier has been adopted within gravity studies performed in the paper \cite{Jaco}. In this case, the above constraint is generalized to a curved space-time as $g^{\mu\nu}u_{\mu}u_{\nu}-1=0$, where $u_{\mu}$ is the aether vector field.

Our starting point is the action \cite{Jaco}
\bea
S=-\frac{1}{16\pi G}\int d^4x\sqrt{-g}\Big[R+\lambda(u^{\mu}u_{\mu}-1)+K^{\alpha\beta}_{\mu\nu}\nabla_{\alpha}u^{\mu}\nabla_{\beta}u^{\nu}
\Big],
\eea
where
\bea
K^{\alpha\beta}_{\mu\nu}=c_1g^{\alpha\beta}g_{\mu\nu}+c_2\delta^{\alpha}_{\mu}\delta^{\beta}_{\nu}+c_3\delta^{\alpha}_{\nu}\delta^{\beta}_{\mu}+c_4u^{\alpha}u^{\beta}g_{\mu\nu}.
\eea
This action involves an above-mentioned constraint introduced with use of the Lagrange multiplier $\lambda$.  The $c_1,c_2,c_3,c_4$ are some dimensionless constants.  It is interesting to note that the term $R_{\alpha\beta}u^{\alpha}u^{\beta}$ proposed as the aether term in \cite{CarTam} arises in this theory (together with some other terms) for the particular case $c_3=-c_2$ when the commutator of covariant derivatives yielding a curvature tensor emerges \cite{Jaco1}.

The corresponding equations of motion look like \cite{Jaco1}:
\bea
&&g_{\alpha\beta}u^{\alpha}u^{\beta}=1; \quad\, \nabla_{\alpha}J^{\alpha}_{\phantom{a}\mu}-c_4\dot{u}_{\alpha}\nabla_{\mu} u^{\alpha}=\lambda u^{\mu};\nonumber\\
T_{\alpha\beta}&=&-\frac{1}{2}g_{\alpha\beta}{\cal L}_u+\nabla_{\mu}\Big(J^{\alpha}_{\phantom{a}(\mu}u_{\beta)}-J^{\mu}_{\phantom{m}(\alpha}u_{\beta)}-J_{(\alpha\beta)}u^{\mu}
\Big)+\\
&+&c_1[(\nabla_{\mu} u_{\alpha})(\nabla^{\mu}u_{\nu})-(\nabla_{\alpha}u_{mu})(\nabla_{\beta}u^{\mu})]+
c_4\dot{u}_{\alpha}\dot{u}_{\beta}+[u_{\nu}\nabla_{\mu} J^{\mu\nu}-c_4\dot{u}^2]u_{\alpha}u_{\beta}.\nonumber
\eea
Here $\dot{u}^{\mu}=u^{\alpha}\nabla_{\alpha}u^{\mu}$, $J^{\alpha}_{\phantom{a}\mu}=K^{\alpha\beta}_{\mu\nu}\nabla_{\beta}u^{\nu}$, and ${\cal L}_u$ is $u$-dependent part of the Lagrangian. We note again that the vector $u_{\mu}$ has nothing to do with the usual matter, so, the Einstein-aether theory is an example of the vector-tensor gravity.

So, now our task will consist in finding some solutions for these equations, or, to be more precise, in checking the consistency of known GR solutions within the Einstein-aether gravity.

As the simplest example we choose the spherically symmetric static metric, which is consistent since the vector $u_{\mu}$ is time-like, in order to satisfy the constraint. In our case, it is convenient to choose this metric in the form slightly different from (\ref{SSSM}), namely,
\bea
ds^2=N(r) dt^2-B(r)(dr^2+r^2d\Omega^2).
\eea
The consistency of this metric within the Einstein-aether gravity has been verified within perturbative methodology for various relations between the parameters $c_1,c_2,c_3$, f.e. $c_1+c_2+c_3=0$, and $c_4$ can be chosen to be zero without any problems since it can be removed through a simple change of variables (see details in \cite{Jaco1}) so that the $N(r)$ and $B(r)$ turn out to be represented as power series in $x=1/r$ providing that they tend to 1 at infinity as it must be, with some lower coefficients in these power series, up to $1/r^3$ terms in large $r$ limit have been explicitly found in certain cases.

For example, treating the black holes solutions, one can show \cite{Jaco1} that the metric
\bea
ds^2=(1-\frac{2M}{r}+\frac{2\beta M^2}{r^2}) dt^2-(1-\frac{2\gamma M}{r})(dr^2+r^2d\Omega^2).
\eea
is consistent in this theory, with $\gamma=1$ (the usual value characteristic for Schwarzschild metric) and $\beta$ expressed in terms of coefficients $c_1,c_2,c_3$. Actually this solution is the Schwarzschild metric modified by the additive term.

Similarly, much more solutions for the Einstein-aether gravity can be obtained, in particular, the cosmological ones. In this context, the detailed study of various cosmological aspects of this theory has been performed in \cite{Paliana} where the model involving two scalar fields coupled to Einstein-aether gravity was considered, and it has been explicitly demonstrated on the base of the numerical analysis of solutions that the consistent potential for these fields is the exponential one, and  the de Sitter-like solutions can arise both in the past (inflationary Universe) and in the future (de Sitter attractor). Earlier the idea of using the Einstein-aether model in order to explain the cosmic acceleration has been claimed in \cite{MengDu}. All this allows to conclude that the Einstein-aether gravity can be considered as an acceptable solution of the dark energy problem. Besides of this, a detailed discussion of various aspects of Einstein-aether gravity, including discussion of plane wave solutions and observational constraints on parameters of the theory, can be found in \cite{JacoRev}. Also, we note that the Einstein-aether gravity also displays some similarity to the Einstein-Maxwell theory, see \cite{Jaco}.

However, it is clear that the Einstein-aether model is problematic from the quantum viewpoint.  Indeed, its action involves a constraint. As it is well known (see f.e. \cite{GN}), a theory with constraints, being considered at the perturbative level, requires special methodologies like $1/N$ expansion which clearly cannot be applied to the Einstein-aether gravity since it involves only four fields $u^{\mu}$. Moreover,  in principle such a theory, when treated in an improper manner, can display various instabilities. Therefore, the natural idea consists in introducing the spontaneous Lorentz symmetry breaking not through constraints but through introducing some potential of the $B_{\mu}$ field displaying a set of minima. This idea gave origin to the bumblebee gravity \cite{KosBluhm,Bertolami} which we begin to discuss now. 

\section{Bumblebee gravity}
 
 So, let us start with considering the bumblebee gravity.  Our initial point will be the action (\ref{bumblebee}). The key features of this action, in comparison with the Einstein-aether theory, are the following ones.
 
 First, this action is characterized by the generic potential, instead of the constraint, which makes it better for quantum studies since the usual perturbative methodology can be applied. Second, the kinetic term is Maxwell-like which is essential to avoid arising of ghost modes \cite{Chkareuli}. Again, the $\pm$ sign reflect the fact that $b^2>0$. We note again that the vacua $B_{0\mu}$ are given by the condition $B^{\mu}_0B_{0\mu}=\pm b^2$, and these vacua are not required to be constants, in a curved space-time, which avoids the difficulties connected with definition of the constant vectors in this case.
 
 First effect to note here is that after Lorentz symmetry breaking, we will have Nambu-Goldstone modes: if we introduce the vector $b_{\mu}$ corresponding to one of the vacua, i.e. $b^{\mu}b_{\mu}=\pm b^2$, define $B_{\mu}=b_{\mu}+A_{\mu}$, and rewrite the action (\ref{bumblebee}) in terms of $b_{\mu}$ and $A_{\mu}$, the resulting form of the action will be given by the Maxwell term, plus the axial gauge term proportional to $(b^{\mu}A_{\mu})^2$, plus new couplings of the vector $A_{\mu}$ with the curvature, like $A^{\mu}A^{\nu}R_{\mu\nu}$, plus the Carroll-like term $b^{\mu}b^{\nu}R_{\mu\nu}$ \cite{CarTam}.
 
 Let us discuss some exact solutions for this theory. First, we consider the static spherically symmetric metric, following the lines of \cite{Bertolami}. For the reasons of convenience, we rewrite the metric (\ref{SSSM}) as:
 \bea
ds^2=-e^{2\phi(r)}dt^2+e^{2\rho(r)}dr^2+r^2d\Omega^2.
\eea
Then, we choose the vacuum vector to be purely radial, i.e. $b_{\mu}=(0,b(r),0,0)$, thus one has $\nabla_{\mu} b_{\nu}=0$ if $b(r)=\xi^{-1/2}b_0e^{\rho(r)}$, $\xi$ is a constant, and the variable $\phi(r)$ becomes irrelevant within modified Einstein equations.

For this metric we find the only non-zero component of the Ricci tensor and the corresponding scalar curvature to be
\bea
R_{rr}=\frac{2\rho^{\prime}}{r};\quad\, R=\frac{2[1+2(r\rho^{\prime}-1)e^{-2\rho}]}{r^2}.
\eea
It is convenient to introduce a new dynamical variable $\Psi=\frac{1-e^{-2\rho}}{r^2}$. Its action will look like:
\bea
S=\frac{2}{\kappa}\int dt dr r^2e^{\rho+\phi}[(3+b^2_0)\Psi+(1+\frac{b^2_0}{2}r\Psi^{\prime})],
\eea
where $b_0$ was defined above.

The equation of motion, after varying with respect to $\phi$, is
\bea
(3+b^2_0)\Psi+(1+\frac{b^2_0}{2}r\Psi^{\prime})=0.
\eea
Its solution is $\Psi(r)=\Psi_0r^{L-3}$, with $3-L=(3+b^2_0)/(1+b^2_0/2)$,
and
\bea
g_{rr}=e^{2\rho}=(1-\Psi_0r^{L-1})^{-1},
\eea
so, this component is similar to $g_{rr}$ of the Schwarzschild metric, therefore our solution is characterized by the event horizon. In principle, more results for this metric can be obtained, f.e. the Hawking temperature \cite{Bertolami}. The case when the $b_{\mu}$ vacuum vector possesses not only the radial component but also the temporal one has been also discussed in \cite{Bertolami}, as a result, the Schwarzschild-like solution will carry extra factor $e^{\pm 2K_ir^{\alpha}}$, where $\alpha$ is a constant, the sign $+$ is for the temporal component, and the sign $-$ for the radial one, with the values of $K_i$ are different for these two components. Therefore, we conclude that the Lorentz symmetry breaking generates the black hole solutions.

Another important example is the cosmological FRW metric. Here we review its description within the bumblebee context presented in \cite{Capelo}.  Explicitly, as a first attempt, we suggest the vector $B_{\mu}$ to be directed along the time axis, $B_{\mu}=(B(t),0,0,0)$. Evidently, in this case the stress tensor for the bumblebee field vanishes, and the only nontrivial component of the equations of motion for the $B_{\mu}$ is
\bea
(V^{\prime}-\frac{3}{2\kappa^2}\frac{\ddot{a}}{a})B=0.
\eea
Thus, the bumblebee field either vanishes or, at $\xi=0$, stays at one of the minima of the potential. In this case, it is possible to show numerically that one has the de Sitter-like expansion of the Universe.

More generic solutions can be obtained for $B_{\mu\nu}\neq 0$. However, in this case the numerical analysis is necessary. Explicit studies carried out in \cite{Capelo} show that in this case, de Sitter-like solutions arise for many values of parameters of the theory confirming this a possibility to have a cosmic acceleration due to the bumblebee field, therefore, one can conclude that the spontaneous Lorentz symmetry breaking can explain the dark energy problem.

Finally, we consider also the G\"{o}del solution (\ref{Godel}). Within the bumblebee context it has been considered in \cite{ourbumb}.
In this case, the energy-momentum tensor is suggested to be a sum of that one for the relativistic fluid (we note that namely this form has been employed in \cite{Godel}):
\bea
\label{tmmn}
T^M_{\mu\nu}=\rho v_{\mu}v_{\nu}+\Lambda g_{\mu\nu},
\eea
and that one for the bumblebee:
\bea
T^B_{\mu\nu}=B_{\mu\alpha}B_{\nu}^{\phantom{\nu}\alpha}-\frac{1}{4}g_{\mu\nu}B_{\lambda\rho}B^{\lambda\rho}-Vg_{\mu\nu}+2V^{\prime}B_{\mu}B_{\nu},
\eea
where $V^{\prime}$ is a derivative of the potential with respect to its argument.
Therefore, the modified Einstein equation (in an appropriate system of units where $\kappa=1$) looks like
\bea
G_{\mu\nu}=T^M_{\mu\nu}+T^B_{\mu\nu}.
\eea
The Einstein tensor $G_{\mu\nu}$ and the matter energy-momentum tensor $T^M_{\mu\nu}$ (\ref{tmmn}) in the bumblebee gravity are the same as in the usual Einstein gravity with the cosmological term. Therefore, the G\"{o}del metric continues to be solution in our theory if and only if the energy-momentum tensor of the bumblebee field will vanish. To achieve this situation, we suggest that the field $B_{\mu}$ is one of the vacua which, for the quartic potential $V=\frac{\lambda}{2}(B^{\mu}B_{\mu}\pm b^2)^2$, will yield vanishing of the potential and its derivative. So, it remains to find the vacuum for which the stress tensor $B_{\mu\nu}=\pa_{\mu}B_{\nu}-\pa_{\nu}B_{\mu}$ would vanish as well (the part proportional to Christoffel symbols vanishes identically). It is clear that the case of the constant $B_{\mu}$ is an excellent example. Some interesting cases of such vacua, for the metric in the form (\ref{Godel}), are: $B_{\mu}=(ab,0,0,0)$, $B_{\mu}=(0,ab,0,0)$, $B_{\mu}=(0,0,0,ab)$ (we note that the G\"{o}del metric is characterized by the constant parameter $a$).
 
 It remains to check consistency of these solutions with the equation of motion for the bumblebee field:
 \bea
 \nabla_{\mu}B^{\mu\nu}=2V^{\prime}(B^2)B^{\nu}.
 \eea  
 These equations are satisfied immediately. Indeed, the l.h.s. is zero since $B^{\mu\nu}=0$ for these solutions, and its covariant derivative is also zero, and the r.h.s. is zero for the quartic potential, if $B_{\mu}$ is one of the vacua. Therefore, we conclude that the 
 G\"{o}del solution is consistent in the bumblebee gravity. More detailed discussion on this solution can be found in \cite{ourbumb}. It is clear that a more generic G\"{o}del-type solution (\ref{gtype}) can be analyzed along the same lines.
 
 An interesting discussion of the bumblebee field is presented also in \cite{Seif}. The starting point is the generalized bumblebee Lagrangian
 \bea
{\cal L}=R-\zeta \bar{g}^{\alpha\gamma}\bar{g}^{\beta\delta}B_{\alpha\beta}B_{\gamma\delta}-V(B^2),
\eea
where $V$ is a some potential of the bumblebee field, $\zeta$ is a coupling constant, and  $\bar{g}^{\alpha\gamma}=g^{\alpha\gamma}+\beta B^{\alpha} B^{\gamma}$ is the effective metric.

Then, we carry out background-quantum splitting for gravitational and bumblebee fields by the formulas $g_{\alpha\beta}=\eta_{\alpha\beta}+h_{\alpha\beta}$ and $B_{\alpha}=\bar{B}_{\alpha}+A_{\alpha}$, where $\bar{B}_{\alpha}$ is one of vacua, i.e. $V(\bar{B}^2)=V^{\prime}(\bar{B}^2)=0$.

As a result, we arrive at the linearized equations of motion for the fluctuations $h_{\alpha\beta}$, $A_{\alpha}$:
\bea
G_{\alpha\beta}|&=&V^{\prime\prime}(\bar{B}^2)\bar{B}_{\alpha}\bar{B}_{\beta}B^2|,\nonumber\\
\bar{\eta}^{\alpha\delta}\bar{\eta}^{\beta\gamma}\pa_{\beta}F_{\gamma\delta}[A]&=&\frac{1}{2\zeta}V^{\prime\prime}(\bar{B}^2)\bar{B}^{\alpha}B^2|.
\eea
where $|$ symbol is for a part linear in fluctuations $h_{\alpha\beta},A_{\alpha}$, f.e. $B^2|=2\bar{B}^{\alpha}A_{\alpha}-\bar{B}^{\alpha}\bar{B}^{\beta}h_{\alpha\beta}$, and $\bar{\eta}^{\alpha\delta}=\eta^{\alpha\delta}+\beta\bar{B}^{\alpha}\bar{B}^{\delta}$ (it is interesting to note that the similar metric arises within aether studies). The $F_{\gamma\delta}[A]=\pa_{\gamma}A_{\delta}-\pa_{\delta} A_{\gamma}$ as usual.

We can introduce background-dependent densities
\bea
\rho_m&=&-V^{\prime\prime}(\bar{B}^2)\bar{B}^2B^2|,\nonumber\\
\rho_e&=&\pm\frac{V^{\prime\prime}(\bar{B}^2)\sqrt{|\bar{B}^2|}}{2\zeta}B^2|
\eea
and a 4-velocity $u_{\alpha}=\pm\frac{\bar{B}_{\alpha}}{\sqrt{|\bar{B}^2|}}$, as a result the equations of motion become
\bea
G_{\alpha\beta}|&=&\rho_m u_{\alpha}u_{\beta},\nonumber\\
\pa^{\beta}F_{\beta\alpha}[A]&=&\rho_e u_{\alpha},
\eea
replaying thus the Einstein and Maxwell equations respectively. 
Effectively we showed that our background field $B_{\mu}$ plays the role of the charged dust. We note that in principle, the $\bar{B}_{\alpha}$ and $A_{\alpha}$ fields can be coupled to usual matter in various manners being treated either as a usual photon or as a some extra particle.

To conclude, we see that the bumblebee gravity can be treated as a sound candidate, first, to implement the Lorentz symmetry breaking within the gravity context, second, to display consistency with astronomical observations, due to validity of most important general relativity solutions. Among other results one can mention study of dispersion relations in a linearized bumblebee gravity where the constant bumblebee field triggers deviations from the standard dispersion relations \cite{FerrMal}. However, much more aspects of the bumblebee gravity, especially problem of validity and consistency of many other solutions, are still to be studied. In this context, one of the most important issues is the study of perturbative aspects of the bumblebee gravity, and only first steps in this study are done now.

\section{Conclusions}

We discussed vector-tensor gravity models.  Just as in the previous chapter, the additional field, in this case the vector one, is treated not as a matter field but as an ingredient of the complete description of the gravity itself. The most important aspect of these models consists in the fact that some of them, namely those ones involving potential terms for the vector field, can be extremely useful within the context of the spontaneous Lorentz symmetry breaking. The known examples of these theories are the Einstein-aether gravity and the bumblebee gravity.
   
The Einstein-aether theory has been formulated earlier. Within it, the potential term generating the spontaneous Lorentz symmetry breaking is implemented through the constraint with the corresponding Lagrange multiplier field.  From one side, this action is rather simple, but from another side, the presence of the constraint generates essential difficulties for the perturbative description.  Therefore, the bumblebee model is certainly much more promising. Moreover, the bumblebee approach displays an advantage in comparison with the naive application of the QFT approach suggesting to couple dynamical fields with the constant vectors (tensors) which, as we already noted, cannot be consistently defined in a curved space-time.
 
The bumblebee approach allows to introduce many Lorentz-breaking vector-tensor terms.  The term $B^{\mu}B^{\nu}R_{\mu\nu}$ from (\ref{bumblebee}) is effectively nothing more that the gravitational aether term proposed in \cite{CarTam}. We note that treating of the $B_{\mu}$ as one of the bumblebee vacua rather than the usual constant vector allows to avoid breaking of the general covariance. In a similar manner, other Lorentz-breaking gravitational terms introduced in \cite{KosGra} can be treated. As a result, relaxing the condition for the Lorentz-breaking vector to be constant, we have a theory consistent with the general covariance requirement.

We note that the term $B^{\mu}B^{\nu}R_{\mu\nu}$ is the particular case of the term $s^{\mu\nu}R_{\mu\nu}$ discussed in \cite{KosGra}. Actually, in \cite{KosGra}, two terms are presented, so, the possible Lorentz-breaking extension of gravity is introduced through adding the term
\bea
\delta S=\int d^4x \sqrt{|g|}(s^{\mu\nu}R_{\mu\nu}+t^{\mu\nu\lambda\rho}R_{\mu\nu\lambda\rho}),
\eea
where $s^{\mu\nu}$, $t^{\mu\nu\lambda\rho}$ are coefficients of explicit Lorentz symmetry breaking (in this review, we consider only the zero torsion case). However, up to now the main attention (see f.e. \cite{FerrMal}) was paid to the $s^{\mu\nu}$ term while the $t^{\mu\nu\lambda\rho}=0$ condition was applied.

To close the discussion of the Lorentz symmetry breaking in gravity, let us say some words about the weak (linearized) gravity. We have noted already that, for the specific form of the Chern-Simons coefficient, the gravitational CS term (\ref{fullCS0}) displays Lorentz symmetry breaking. In \cite{ourlingrav}, another, one-derivative Lorentz-breaking term in the linearized gravity has been studied. In principle, much more Lorentz-breaking terms in the linearized gravity can be introduced. However, it is clear that many studies of Lorentz symmetry breaking in gravity are still to be carried out, and it is natural to expect that such studies will be performed in the next years.
   
\chapter{Horava-Lifshitz gravity}   

\section{Introduction}

As it is well known, the most complicated problem of the gravity is the problem of its consistent quantum description. Indeed, we have noted in the Chapter 1 that the Einstein gravity is non-renormalizable since the mass dimension of the gravitational constant is negative. The natural improvement of situation could consist in adding the higher-derivative terms which clearly make the UV asymptotics of the propagator better. However, it is known that in this case the ghosts arise which makes the theory to be unstable, hence higher-derivative gravity models can be used only as effective theories for the low-energy domain. 
   
Therefore, in \cite{Horava}, the following idea has been proposed: let us suggest that the desired extension of gravity involves only second time derivatives, so, the ghosts will be ruled out, and higher spatial derivatives, therefore the UV behavior of the propagator will be improved. The similar models for the scalar field, with modified kinetic terms like $\frac{1}{2}\phi(\partial^2_0+(-1)^z\alpha \Delta^z)\phi$ have been introduced a long ago within the condensed matter context in \cite{Lifshitz} where they were used to describe critical phenomena. In other words, we suggest that the Lorentz symmetry breaking is strong. Further, such theories with strong difference between spatial and time directions have been denominated as theories with space-time anisotropy. The number $z$, defined in a manner similar to the action above (once more, if the action involves two time derivatives, it involves $2z$ spatial derivatives), is called the critical exponent. For the Lorentz-invariant theories, one has $z=1$.  To recover the Einstein limit, one must suppose that the action involves also lower-derivative terms. One can verify that in such a theory, the dimension of the effective gravitational constant will depend on $z$, being actually equal to $z-d$, in a $d$-dimensional space-time. Therefore, in $(3+1)$-dimensional space-time, the gravity model formulated on the base of the space-time anisotropy (further such theories became to be called the Horava-Lifshitz (HL) theories) is power-counting renormalizable at $z=3$. However, it is clear that for such a theory, the perturbative calculations will be very involved.

In this chapter we present a general review on HL gravity, introduce definitions of main quantities used within it, and describe most important classical solutions.

\section{Basic definitions}
 
 So, let us construct the gravity model on the base of a strong difference between time and space coordinates. Following the methodology developed in \cite{Horava}, we consider the space-time as a foliation $R\times M_3$, where $R$ is the real axis corresponding to the time, and $M_3$ is the three-dimensional manifold parametrized by spatial coordinates.  The most convenient variables to parametrize the gravitational field in this case are the Arnowitt-Deser-Misner (ADM) variables \cite{ADM}, that is, $N,N_i,g_{ij}$ defined from the following representation of the metric:
 \begin{eqnarray}
ds^2&=&g_{\mu\nu}dx^{\mu}dx^{\nu}\equiv g_{00}\,dt^2 + 2\,g_{0i}\, dx^idt + g_{ij}\, dx^idx^j=\nonumber\\
&=&-N^2dt^2+g_{ij}(dx^i+N^idt)(dx^j+N^jdt),
\end{eqnarray}
so, $g_{ij}$ is the purely spatial metric, and one has the shift vector $N_i=g_{0i}$ and the lapse function $N=(g_{ij}N^iN^j-g_{00})^{1/2}$.

The Lagrangian was suggested to be in the form
\begin{eqnarray} \label{lagra}
L&=&\sqrt{g}N\Big(\frac{2}{\kappa^2}(K_{ij}K^{ij}-\lambda K^2)-\frac{\kappa^2}{2w^4}C_{ij}C^{ij}
+\frac{\kappa^2\mu}{2w^2}\frac{\epsilon^{ijk}}{\sqrt{g}}R_{il}\nabla_jR^l_k
-\nonumber\\ &-&
\frac{\kappa^2\mu^2}{8}R_{ij}R^{ij} 
+\frac{\kappa^2\mu^2}{8(1-3\lambda)}[\frac{1-4\lambda}{4}R^2+\Lambda R-3\Lambda^2]
+{\cal L}_m\Big), 
\end{eqnarray}
where the $R_{ij}$ is a purely spatial curvature constructed on the base of the spatial metric $g_{ij}$, and
\bea
\label{extcurv}
K_{ij}=\frac{1}{2N}(\dot{g}_{ij}-\nabla_i N_j-\nabla_j N_i),
\eea
is the extrinsic curvature, with the dot is for a derivative with respect to $t$, $K=g^{ij}K_{ij}$, and
\bea
C^{ij}=\frac{\epsilon^{ikl}}{\sqrt{g}}\nabla_k(R^j_l-\frac{1}{4}R\delta^j_l)
\eea
is a Cotton tensor. It involves three spatial derivatives, hence the term $C_{ij}C^{ij}$ is of the sixth order. So, it is clear that the propagator in this theory behaves as $G(k)\sim\frac{1}{k^2_0-\vec{k}^6}$. As we already noted, this implies power-counting renormalizability of the theory, and the gravitational constant $\kappa$ is indeed dimensionless.

The form of the Lagrangian (\ref{lagra}) has been motivated by "detailed balance" condition \cite{Horava}  requiring that the potential term (i.e. the part of the action which does not involve the extrinsic curvature $K_{ij}$ which only includes the time derivatives) is
\bea
\label{detbal}
S_V=\frac{\kappa^2}{8}\sqrt{g}N\frac{\delta W}{\delta g_{ij}}G_{ijkl}\frac{\delta W}{\delta g_{kl}},
\eea
where $W$ is a some action, and $G_{ijkl}=\frac{1}{2}(g_{ik}g_{jl}+g_{il}g_{jk}-\lambda g_{ij}g_{kl})$.
For $z=2$, one has $W=W_2=\frac{1}{2\kappa_W}\int d^Dx \sqrt{g}(R-2\Lambda_W)$, and for $z=3$, one chooses $W=W_3$ to be the $3D$ Chern-Simons action, so, $\frac{\delta W_3}{\delta g_{ij}}=C^{ij}$ (a similar expression for the Cotton tensor in $2+1$ dimensions has been considered in the section 3.2), and substitution of $W=W_2+W_3$ to (\ref{detbal}) yields the potential term given by (\ref{lagra}).

However, there are only very few attempts to do quantum calculations in the HL gravity \cite{Mazzi}. Actually, in these papers the gravity is suggested to be a pure background field, only the matter is quantized. At the same time, it is clear that the calculations of quantum corrections in a pure HL gravity, besides being extremely involved technically, must answer the fundamental question -- whether the form of  quantum corrections matches the form of the classical action, i.e. whether the HL gravity is multiplicatively renormalizable? This question is still open.

Let us now write down the equations of motion for the HL gravity. We use approach and notations from \cite{KirKof} with $Q_{kl}=N(\gamma R_{kl}+2\beta C_{kl})$.
It should be noted that $g_{00}$ is not a fundamental dynamical variable of the theory. For $g_{00}$ one has
\bea
\label{eqG00}
\frac{\delta S}{\delta g_{00}} &=&
(\frac{\delta S_{g}}{\delta N}+\frac{\delta S_{m}}{\delta N})\frac{\delta N}{\delta g_{00}}   = G^{00}-T^{00}=0.
\eea 
We note that, since $N=(g_{ij}N^iN^j-g_{00})^{1/2}$, one has $\frac{\delta N}{\delta g_{00}}=-\frac{1}{2N}$.
Hence,
\bea
\label{tens1}
G^{00}&=&\frac{1}{2N}(-\alpha(K_{ij}K^{ij}-\lambda K^2)+\beta C_{ij}C^{ij} + \sigma \\  \nonumber
&+&\gamma\frac{\epsilon^{ijk}}{\sqrt{g}}R_{il}\nabla_jR^l_k
+\zeta R_{ij}R^{ij}+\eta R^2+\xi R),  
\eea
where
\bea
\label{def}
&&\alpha=\frac{2}{\kappa^2}\,,\quad\,\beta=-\frac{\kappa^2}{2w^4}\,,
\quad\,\gamma=\frac{\kappa^2\mu}{2w^2}\,,\quad\,\zeta=-\frac{\kappa^2\mu^2}{8}; \nonumber\\
\nonumber \\
&&\eta=\frac{\kappa^2\mu^2(1-4\lambda)}{32(1-3\lambda)},
\quad \xi=\frac{\kappa^2\mu^2\Lambda}{8(1-3\lambda)}\,, \nonumber \\
&&
\sigma=-\frac{3\kappa^2\mu^2\Lambda^2}{8(1-3\lambda)}
\eea
are constant parameters of the theory.

For $g_{0i}$ we find
\bea
\label{eqG0i}
\frac{\delta S}{\delta g_{0l}}=\frac{\delta S}{\delta N_l}=2\alpha\nabla_k(K^{kl}-\lambda Kg^{kl})-T^{0l}=0.
\eea

Finally, for $g_{ij}$ we find
\bea
\label{eqGij}
G_{ij}=T_{ij},
\eea
where 
\bea
\label{einh}
G_{ij}=G^{(1)}_{ij}+G^{(2)}_{ij}+G^{(3)}_{ij}+G^{(4)}_{ij}+G^{(5)}_{ij}+G^{(6)}_{ij}.
\eea  

Here, with $\Box\equiv\nabla^2$, one has
\bea
\label{eincomp}
G^{(1)}_{ij}&=&2\alpha NK_{ik}K_j^k-\frac{\alpha N}{2}K_{kl}K^{kl}g_{ij}+ 
\alpha(K_{ik}N_j)^{;k}+\alpha(K_{jk}N_i)^{;k}-\nonumber\\ &-&\alpha(K_{ij}N_k)^{;k}+(i\leftrightarrow j)\,,
\nonumber\\
G^{(2)}_{ij}&=&-2\alpha \lambda NKK_{ij}+\frac{\alpha\lambda
N}{2}K^2g_{ij}-\frac{\alpha\lambda}{\sqrt{g}}g_{ik}g_{jl}\frac{\partial}{\partial t}(\sqrt{g}Kg^{kl})
 \nonumber \\&-&
\alpha\lambda(Kg_{ik}N_j)^{;k}-\alpha\lambda(Kg_{jk}N_i)^{;k}+\alpha\lambda(Kg_{ij}N_k)^{;k}+(i\leftrightarrow j)\,,
\nonumber\\
G^{(3)}_{ij}&=&N\xi R_{ij}-\frac{N}{2}(\xi R+\sigma)g_{ij}-\xi N_{;ij}+\xi \Box N g_{ij}+(i\leftrightarrow j)\,,
\nonumber\\
G^{(4)}_{ij}&=&2N\eta RR_{ij}-\frac{N}{2}\eta R^2g_{ij}+2\eta \Box(N R)g_{ij}-2\eta(NR)_{;ij}+(i\leftrightarrow j)\,,
\nonumber\\
G^{(5)}_{ij}&=&\Box (N (\zeta R_{ij}+\frac{\gamma}{2}C_{ij})) -(N(\zeta R_{ki}
+\frac{\gamma}{2}C_{ki}))_{;j}^{;\phantom{j}k}+\nonumber\\&+&
 (N (\zeta R^{kl}+\frac{\gamma}{2}C^{kl}))_{;lk}\,g_{ij}+ (i\leftrightarrow j)\,,
 \nonumber
\eea
and
\bea
G^{(6)}_{ij}&=&\frac{1}{2}\frac{\epsilon^{mkl}}{\sqrt{g}}\Big[(Q_{mi})_{;kjl}+(Q_m^{\phantom{k}n})_{;kin}g_{jl}-\nonumber\\ &-&
(Q_{mi})_{;kn}^{;\phantom{kk}n}g_{jl}-(Q_{mi})_{;k}R_{jl}-
(Q_{mi}R_k^n)_{;n}g_{jl}  \nonumber\\&+&
(Q_{\phantom{n}m}^nR_{ki})_{;n}g_{jl}+\frac{1}{2}(R^n_{\phantom{n}pkl}Q^{\phantom{p}p}_m)_{;n}g_{ij}+Q_{mi}R_{jl;k}
\Big]+\nonumber\\ &+& 2N\zeta R_{ik}R_j^k  \nonumber \\&-&
\frac{N}{2}(\beta C_{kl}C^{kl}+\gamma R_{kl}C^{kl}+\zeta R_{kl}R^{kl})g_{ij}-\frac{1}{2}Q_{kl}C^{kl}g_{ij}+\nonumber\\
&+&(i\leftrightarrow j)\;.
\eea
These equations are very involved. However, already now we can indicate some situations where the equations are essentially simplified.

First of all, it is clear that the equations of motion are simplified when the variable to be found depends only on one argument (examples of such variables are scale factor and radial function). Second, the case of a diagonal metric simplifies the system immediately since one has $N_i=0$ and $N=\sqrt{|g_{00}|}$. We note that the static spherically symmetric metric (f.e. non-rotating black hole) and FRW metric are diagonal. As for the G\"{o}del metric, it has been considered in a tetrad base which strongly simplifies calculations (see \cite{Rebour} for details). Now, let us consider examples of solutions for gravitational field equations.

\section{Exact solutions}

So, let us consider the exact solutions. Again, as earlier, we consider three examples -- cosmological FRW metric, black hole and 
G\"{o}del-type metric.

We follow \cite{KirKof}. So, for the cosmological case, one suggests $N=N(t)$, $N_i=0$ (since the FRW metric is diagonal), and $g_{ij}=a^2(t)\gamma_{ij}$, where $\gamma_{ij}$ is the maximally symmetric spatial metric yielding constant scalar curvature: $R=6k$, and $R_{ij}=2k\gamma_{ij}$, therefore $\nabla_i R=0$, and the Cotton tensor is also zero, $C_{ij}=0$.  The matter is suggested to be the function of time only, $\Phi=\Phi(t)$. We can introduce the new Hubble parameter $H=\frac{\dot{a}}{Na}$, where $a=a(t)$ is the usual scale factor in (\ref{FRW}).

It is natural to suggest that the matter is given by a scalar field which, as usual in cosmology, depends only on time.
As a result, the equation of motion for $N$ looks like:
\bea
3\alpha(3\lambda-1)H^2+\sigma+\frac{6k\xi}{a^2}+\frac{12k^2(\zeta+3\eta)}{a^4}=\frac{\dot{\Phi}^2}{N^2}+V(\Phi).
\eea
For $g_{ij}$, one finds
\bea
&&2\alpha(3\lambda-1)(\dot{H}+\frac{3}{2}H^2)+\sigma+\frac{2k\xi}{a^2}-\frac{4k^2(\zeta+3\eta)}{a^4}
=\nonumber\\
&=& -\frac{\dot{\Phi}^2}{N^2}+V(\Phi).
\eea
Finally, for a matter the equation is
\bea
\frac{1}{N}\partial_t(\frac{\dot{\Phi}}{N})+3H\frac{\dot{\Phi}}{N}+\frac{1}{2}V_{\Phi}=0.
\eea
One can verify that cyclic or bouncing solutions are possible \cite{Mukohy}. In the vacuum case one can prove directly the possibility of static solutions, while in the presence of the matter, the solutions can be obtained only numerically \cite{LuPope}.

We can have also static spherically symmetric solutions described by the Eq. (\ref{SSSM}). Clearly, the possibility of black holes is of the special interest. We start with the particular case of the metric (\ref{SSSM}):
\bea
ds^2=-f(r)dt^2+\frac{dr^2}{f(r)}+r^2d\Omega^2,
\eea
It is clear that the Schwarzschild and Reissner-Nordstrom metrics match this form.

In \cite{Kena} it has been explicitly shown that, for $\lambda=1$, one has 
\bea
f(r)=1+\omega r^2-\sqrt{r(\omega^2r^3+4\omega M)},
\eea
The $\omega$ is a function of constant parameters of the theory.
The essential conclusion is that at large distances, i.e. $r\ll (M/\omega)^{1/3}$, one has $f(r)\simeq 1-\frac{2M}{r}+O(r^{-4})$, that is, the Schwarzschild result, i.e. the consistency with the general relativity is achieved.

It has been demonstrated in \cite{Kena} that the equation $f(r)=0$ has two solutions, so this black hole has two horizons with $r_{pm}=M(1\pm\sqrt{1-\frac{1}{2\omega M^2}})$. The naked singularity is avoided at $\omega M^2\geq 1/2$.

Now, let us consider the G\"{o}del-type solution (\ref{gtype}). It has been considered in details in \cite{Rebour}. First of all, we note that
$g_{\phi\phi}=D^2-H^2=G(r)$ (other two components of $g_{ij}$ are 1), and $N=\frac{D(r)}{\sqrt{G(r)}}$. So, the positiveness of $G(r)$, and hence satisfying the causality condition, is necessary to have a consistent (real) value of $N$!

After some change of variables discussed in \cite{Rebour}, we can rewrite this metric as
\bea
ds^2=-(dt'+\frac{2\omega}{m}e^{mx}dy)^2+e^{2mx}dy^2+dr^2+{dz'}^2,
\eea
with $G(x)=v^2e^{2mx}>0$, and $v^2=1-\frac{4\omega^2}{m^2}$, so, the causality is guaranteed if $v^2>0$.
For this metric, $R_{1212}=-m^2v^2e^{2mx}$, $K_{12}=-v\omega e^{mx}$, $C^{ij}=0$, $R=-2m^2$.

To verify the consistency of this solution, we choose the fluid-like matter with
\bea
T^{\mu\nu}=(p+\rho) u^{\mu}u^{\nu}+p g^{\mu\nu}.
\eea
Namely this matter has been used in the original paper \cite{Godel}.
After solving algebraic equations we find $m^2=\frac{2}{3}\omega^2$ or $m^2=\frac{1}{4}\omega^2$. However, both these solutions appear to be not completely satisfactory since they are non-causal (as it has been proved in \cite{RebTio}, the causality is achieved for $m^2\geq 4\omega^2)$. As for constant parameters of the theory $\lambda$, $\mu$, $\Lambda$, they  can also be found in terms of $m$, $\omega$, $p$, $\rho$, the explicit values are given in \cite{Rebour}.

Therefore, we have seen that these solutions of GR are consistent within the HL gravity, at least asymptotically. Again, it is important to note that the HL gravity is power-counting renormalizable (although up to now there is no examples of full-fledged quantum calculations in the theory). Nevertheless, it must be noted that it also displays some difficulties which we will discuss now.

\section{Modified versions of HL gravity}

While the HL gravity seems to solve the problem of renormalizability, and the most important classical solutions in it reproduce those ones for the GR in certain limits, the consistent description of degrees of freedom in HL gravity turns out to be problematic. This fact has been firstly described in \cite{Sib1}. Following that paper, the main problem of the HL gravity is as follows: the full-fledged general covariance group is broken up to the subgroup which leaves the space-time foliation to be invariant. In other words, since there is no more symmetry between space and time, one has the reduced gauge group for spatial coordinates only. Thus, the gauge symmetry is partially broken, which implies in arising of new degrees of freedom which can imply unstable vacuum, strong coupling and other unusual effects \cite{Charmousis}. It was claimed in \cite{Sib1} that, actually, the extra mode appears to satisfy the first-order equation of motion and hence does not propagate. 

To illustrate this fact, let us consider the equations of motion (\ref{eqG00},\ref{eqG0i},\ref{eqGij}). As we already noted, they are invariant under three-dimensional gauge transformations in linearized case looking like $\delta g_{ij}=\pa_i\xi_j+\pa_j\xi_i$. These transformations allow to impose the gauge $N_i=0$ \cite{Sib1,Charmousis}. Afterwards, the Eq. (\ref{extcurv}) takes the form: $\dot{g}_{ij}=2NK_{ij}$. However, in the system (\ref{eqG00},\ref{eqG0i},\ref{eqGij}) there is no equation for the evolution of $N$! And since $N$ is separated from all other dynamical variables, it cannot be fixed by gauge transformations. As a result, one concludes that $N$ describes the new degree of freedom. To study it we take the time derivative of (\ref{tens1}),  combine it with other equations, and arrive at
\bea
\label{eqthree}
\nabla_i\left(N^2\Big[\xi(\lambda-1)\nabla^iK+F^i(K_{jk},R_{jk},K)
\Big]
\right)=0.
\eea
It is easy to see that we have 13 dynamical variables ($K_{ij},g_{ij},N$), five constraints given by (\ref{eqG00},\ref{eqG0i},\ref{eqthree}), so, we rest with 8 independent variables. Using three gauge parameters $\xi_i$ we can eliminate three variables more. For five remaining ones, we have four initial conditions for two helicities of $h_{ij}$. So, we stay with one extra degree of freedom!

More detailed analysis performed in \cite{Sib1} shows that if we consider $k_{ij}$, a small fluctuation of $K_{ij}$, its trace $\kappa=k_i^i$ does not propagate since $\nabla^2\kappa=0$. So, we can conclude that this extra mode is non-physical.

Returning to dynamics of $N$, we can fix $N$ through the additive term in the action given by $S_n=\int d^3xdt\sqrt{g}N\frac{\rho}{2}(N^{-2}-1)$, which implies strong coupling (roughly speaking, due to the presence of the constraint). Under some tricks like covariant extension (i.e. introducing of a Lorentz-covariant analogue), it appears to be equivalent to the Einstein-aether action (with $\phi$ is a Stuckelberg field) $S_n=\int d^3xdt\sqrt{g}\frac{\rho}{2}(\nabla_{\mu}\phi\nabla^{\mu}\phi-1)$ \cite{Sib2}, $\phi$ is called chronon since there is a gauge in which this field is equal to a time coordinate, $\phi=t$. 

It was argued in \cite{Sib2} that if we introduce $u_{\mu}=\frac{\pa_{\mu}\phi}{\sqrt{X}}$, with $X=g_{\mu\nu}\pa^{\mu}\phi\pa^{\nu}\phi$, we can add some terms to our action to get a consistent theory! Actually, we have
\bea
S&=&-\frac{1}{\kappa^2}\int d^4x\sqrt{-g}(R_4+(\lambda-1)(\nabla_{\mu}u^{\mu})^2+
\alpha u^{\mu}(\nabla_{\mu}u^{\nu})u^{\lambda}(\nabla_{\lambda}u_{\nu})+\ldots),
\eea
and this action, for splitting $\phi\to t+\chi$, yields reasonable dispersion relations for $\chi$ like $\omega^2=C \vec{p}^2$, with $C$ is a some number. In \cite{Sib2}, also some cosmological impacts of this term were studied. An aside result is an emergence of Einstein-aether action. So, the consistent extension of the HL gravity is found.

Another approach is based on use of so-called projectable version of the HL gravity, where the lapse $N$ is suggested to be a function of a time only, $N=N(t)$. However, it turns out to be that although in this case the theory is strongly simplified, the scalar excitation is still unstable and cannot be ruled out \cite{Sot}.

\section{Conclusions}

Let us make some conclusions regarding the HL gravity. As we already noted, the key idea of the HL gravity is that the usual general covariance is an essentially low-energy phenomenon but not a fundamental feature of the nature.  In a certain sense, it can be said that the HL concept was developed to "sacrifice" general covariance in order to conciliate desired renormalizability with absence of ghosts. In this context, it should be noted that breaking of general covariance in gravity is discussed as well in "usual" Lorentz-breaking gravity models without strong space-time asymmetry \cite{KosMew}. 

We demonstrated how the known GR solutions are modified within the HL context. Within the cosmological context, accelerated and bouncing solutions are possible, thus the HL gravity is a good candidate to solve the dark energy problem. We demonstrated that there are black hole solutions behaving like usual Schwarzschild BHs at large distances. Also, we demonstrated that the G\"{o}del-type solutions consistent within the HL gravity are non-causal, but one should note that the G\"{o}del solution itself is non-causal.

However, quantum description of the HL gravity is rather problematic. One of the reasons is a very complicated structure of the classical action potentially implying a very large number of divergent contributions, therefore while the HL is power counting renormalizable, we cannot yet be sure that it is multiplicatively renormalizable. Another difficulty is the question about an extra degree of freedom. While it was in principle solved in \cite{Sib2}, where the "healthy extension" of HL gravity was introduced, the problem now consists in obtaining physically consistent results on the base of this extension. Therefore, even in this case we have more questions than answers.
To close the discussion, we recommend an excellent review on HL gravity presented in \cite{Wang}.
 
\chapter{Nonlocal gravity} 

\section{Motivations}

As we have noted several times along this review, the main problem of various gravity models is the development of a consistent quantum description. Indeed, the Einstein gravity is  non-renormalizable, and introduction of higher-derivative additive terms implies in arising of ghosts. We have argued in the previous chapter that the Horava-Lifshitz gravity seems to be a good solution since it is power-counting renormalizable, and ghosts ate absent since the action involves only second time derivatives. However, the HL gravity, first, is very complicated, second, breaks the Lorentz symmetry strongly, third, displays a problem of extra degrees of freedom whose solving, as we noted, requires special efforts. At the same time, the concept of nonlocality developed originally within phenomenological context  in order to describe finite-size effects (see f.e. \cite{Efimov}), began to attract  the interest. Besides of this, the nonlocality enjoys also a stringy motivation since the factors like $e^{\Box}$ emerge naturally within the string context \cite{SiegelNL}. The key idea of nonlocal field theories looks like follows. Let us consider for example the free scalar field whose Lagrangian is
\bea
{\cal L}=\frac{1}{2}\phi f(\Box/\Lambda^2)\phi,
\eea
where $f(z)$ is a some non-polynomial function (with $\Lambda$ is the characteristic nonlocality scale) which we choose to satisfy the following requirements.

First, at small arguments this function should behave as $f(z)=a+z$, in order to provide the correct $\Box+m^2$ IR asymptotic behavior. Second, this function must decay rapidly at $|z|\to\infty$ (in principle, we can consider only Euclidean space, so, $z$ is essentially positive), so that integrals like $\int_0^{\infty}f(z)z^ndz$ are finite for any finite non-negative $n$, to guarantee finiteness of the theory (in principle in some case this requirement is weakened, if the theory is required to be not finite but only renormalizable). Third, the $f(z)$ is required to be so-called entire function, i.e. it cannot be presented in the form of a product of primitive multipliers like $(z-a_1)(z-a_2)\ldots$, so, its propagator has no different poles (as we noted in the Chapter 2, namely presence of such a set of poles implies in existence of ghost modes). The simplest example of such a function is the exponential, $f(z)=e^{-z}$. 

Another motivations for nonlocality are the loop quantum gravity dealing with finite-size objects, and the noncommutativity, where the Moyal product is essentially nonlocal by construction. At the same time, it is interesting to note that although the so-called coherent states approach \cite{Spall} has been motivated by quantum mechanics, by its essence it represents itself as  a natural manner to implement nonlocality, so that all propagators carry the factor $e^{-\theta k^2}$, with $\theta$ is the noncommutativity parameter. Within the gravity context, use of the nonlocal methodology appears to be especially promising since it is expected that the nonlocality, being implemented in a proper manner, can allow to achieve renormalizability without paying the price of arising the ghosts. The first step in this study has been done in the seminal paper \cite{Modesto}.

\section{Some results in non-gravitational nonlocal theories}

Before embarking to studies of gravity, let us first discuss the most interesting results in non-gravitational nonlocal theories, especially within the context of quantum corrections.

As we already noted, effectively the nonlocal methodology has been applied to perturbative studies for the first time within the coherent states approach \cite{Spall} which includes Gaussian propagator guaranteeing convergence of quantum corrections. Further, various other studies have been performed. An important role was played by the paper \cite{Briscese} where the effective potential in a nonlocal theory has been calculated for the first time. In that paper, the following theory has been introduced:
\bea
{\cal L}=-\frac{1}{2}\phi(\exp(\Box/\Lambda^2)\Box+m^2)\phi-V(\phi).
\eea
Here, $\Lambda$ is a characteristic nonlocality scale.
For this theory, one can calculate the one-loop effective potential given by the following integral:
\bea
\label{epnl}
V^{(1)}=\frac{1}{2}\int\frac{d^4k_E}{(2\pi)^4}\ln\left( \exp(-\frac{k^2_E}{\Lambda^2}) k^2_E+m^2+V^{\prime\prime}\right). 
\eea
 It is clear that at $k^2\ll \Lambda^2$, the theory is reduced to usual one. 
The exponential factors guarantee finiteness. It is easy to see that there is no ghosts in the theory since there is no different denominators $\Box+m^2_i$ in the propagator of the theory. However, the integral (\ref{epnl}) can be calculated only approximately for various limits, and it is easy to see that it diverges as $\Lambda\to\infty$ (in \cite{Briscese}, a some procedure to isolate this divergence has been adopted). Further, this study has been generalized for the superfield theories representing themselves as various nonlocal extensions of Wess-Zumino model and super-QED, in \cite{ournloc}. It is clear that when, in these theories, one consider the limit of an infinite nonlocality scale $\Lambda\to \infty$, the theory returns to the local limit and becomes to be divergent, i.e. the nonlocality acts as a kind of the higher-derivative regularization, so, the quantum contributions are singular in this limit growing as $\Lambda^2$ if the local counterpart of the theory involves quadratic divergences, or as $\ln \Lambda^2$, if it involves the logarithmic ones. From a formal viewpoint, the existence of these singularity can be exemplified by the fact that the typical integral in nonlocal (Euclidean) theory grows quadratically with $\Lambda$ scale since $\int\frac{d^4k}{(2\pi)^4}\frac{1}{k^2}e^{-k^2/\Lambda^2}\propto \Lambda^2$. Effectively, the problem of the singularity of the result at $\Lambda\to\infty$ is nothing more that the problem of large quantum corrections arising also in higher-derivative and noncommutative field theories.

At the same time, the problems of unitarity and causality in nonlocal theories require special attention since the nonlocality is commonly associated with an instant propagation of a signal. These problems were discussed in details in various papers. So, it has been claimed in \cite{Tomboulis} that the problems of unitarity and causality can be solved at least for certain forms of nonlocal functions. Further this result was corroborated and discussed in more details in \cite{Bric2}. However, the complete discussion of unitarity and causality in nonlocal field theories is still to be done. Otherwise, the nonlocal theories must be treated only as effective ones.

So, to go to studies of gravity, we can formulate some preliminary conclusions: (i) there is a mechanism allowing to avoid UV divergences: (ii) this mechanism is Lorentz covariant and ghost free: (iii) the unitarity and causality still are to be studied.

\section{Classical solutions in nonlocal gravity models}

So, let us introduce examples of nonlocal gravity models. The paradigmatic example has been proposed in \cite{Bisw1}, where the Lagrangian
${\cal L}=\frac{1}{G}\sqrt{|g|}F(R)$ was studied, with
\bea
\label{firstact}
F(R)=R-\frac{R}{6}(\frac{e^{-\Box/M^2}-1}{\Box})R.
\eea
Here the d'Alembertian operator $\Box$ is covariant: $\Box=g^{\mu\nu}\nabla_{\mu}\nabla_{\nu}$. This is the nonlocal extension of $R^2$-gravity. 

First of all, it is easy to show that this theory is ghost-free. Indeed, we can expand
\bea
F(R)=R+\sum\limits_{n=0}^{\infty}\frac{c_n}{M^{2n+2}}R\Box^n R,
\eea
with $c_n=-\frac{1}{6}\frac{(-1)^{n+1}}{(n+1)!}$.
We can rewrite this Lagrangian with auxiliary field $\Phi$ and scalar $\psi$ (we can eliminate first $\Phi$, and then $\psi$, through their equations of motion):
\bea
{\cal L}=\frac{1}{G}\sqrt{|g|}(\Phi R+\psi\sum\limits_{n=1}^{\infty}\frac{c_n}{M^{2n+2}}\Box^n\psi-[\psi(\Phi-1)-\frac{c_0}{M^2}\psi^2] ).
\eea
Then we do conformal transformations $g_{mn}\to \Phi g_{mn}$, with $\Phi\simeq 1+\phi$, to absorb $\Phi$ in curvature term.
As a result, we arrive at the Lagrangian
\bea
{\cal L}=\frac{1}{G}\sqrt{|g|}( R+\psi\sum\limits_{n=0}^{\infty}\frac{c_n}{M^{2n+2}}\Box^n\psi-\psi\phi+\frac{3}{2}\phi\Box\phi ).
\eea
with the equations of motion are
\bea
\psi=3\Box\phi;\quad\, \phi=2\sum\limits_{n=0}^{\infty}\frac{c_n}{M^{2n+2}}\Box^n\psi.
\eea
From here we have equation of motion for $\phi$:
\bea
(1-6\sum\limits_{n=1}^{\infty}c_n\frac{\Box^{n+1}}{M^{2n+2}})\phi=[1+\frac{e^{\Box/M^2}-1}{\Box/M^2}]\phi=0,
\eea
The l.h.s. is evidently entire, so we have no ghosts.

We conclude that the nonlocality in gravity sector can be transferred to matter sector! This is valid for various models. In a certain sense, this fact is analogous to the observation made in the section 2.3 where it was argued that the $f(R)$ gravity, representing itself as an example of higher-derivative theory, can be mapped to a some scalar-tensor gravity with no higher derivatives in the gravity sector.

The Lagrangian (\ref{firstact}) can be rewritten as \cite{Bisw2}:
\bea
{\cal L}=\sqrt{|g|}\Big(\frac{1}{G}R+\frac{\lambda}{2}RF(\Box)R-\Lambda+{\cal L}_M 
\Big).
\eea
The function $F(\Box)$ is assumed to be analytic, as it is motivated by string theory, and, moreover, in the analytic case the theory does not display problems in IR limit. The Gaussian case, which is especially convenient from the viewpoint of the UV finiteness, is the perfect example.
The equations of motion, for $M^2_P=G^{-1}$, take the form
\bea
\label{cosmeq}
[\frac{M^2_P}{2}&+&2\lambda F(\Box)R]G^{\mu}_{\nu}=T^{\mu}_{\nu}+\Lambda\delta^{\mu}_{\nu}+\lambda K^{\mu}_{\nu}-
\frac{\lambda}{2}(K^{\alpha}_{\alpha}+K_1)-\nonumber\\ &-&\frac{\lambda}{2}RF(\Box)R\delta^{\mu}_{\nu}
+2\lambda(g^{\mu\alpha}\nabla_{\alpha}\nabla_{\nu}-\delta^{\mu}_{\nu}\Box)F(\Box)R,\\
K^{\mu}_{\nu}&=&g^{\mu\rho}\sum_{n=1}^{\infty}f_n\sum_{l=0}^{n-1}\pa_{\rho}\Box^lR\,\pa_{\mu}\Box^{n-l-1}R;\nonumber\\
K_1&=&\sum_{n=1}^{\infty}f_n\sum_{l=0}^{n-1}\Box^l R\, \Box^{n-l}R;\quad\, F(\Box)=\sum\limits_{n=0}^{\infty}f_n\Box^n.\nonumber
\eea
It is important to note that in two last lines $\Box^l$ acts {\bf only} to the adjacent $R$.

Now, the natural problem is finding some solutions of these equations. In \cite{Bisw2}, the following ansatz has been proposed,
with $r_1,r_2$ are some real numbers:
\bea
\label{ansatz}
\Box R-r_1R-r_2=0
\eea
which implies (here $f_0$ is zeroth order in expansion of $F(\Box)$ in series)
\bea
F(\Box)R=F(r_1)R+\frac{r_2}{r_1}(F(r_1)-f_0).
\eea
This allows to reduce the order of equations to at maximum second. It is clear that constant curvature makes the equation trivial, just this situation occurs for G\"{o}del-type solutions.

One can found nontrivial cosmological solutions for this theory. In particular, bouncing solutions, for $r_1>0$, are possible:
\bea
\label{cosh}
a(t)=a_0\cosh(\sqrt{\frac{r_1}{2}}t).
\eea
Let us give more details for cosmology. Indeed, if we substitute the FRW metric (\ref{FRW}) to (\ref{cosmeq}), and suggest that, as usual in cosmology, $\rho=\rho_0(\frac{a_0}{a})^4$, we have from (\ref{ansatz}), with $r_1\neq 0$:
\bea
\label{solh}
\frac{d^3H}{dt^3}+7H\ddot{H}+4\dot{H}^2-12H^2\dot{H}=-2r_1H^2-r_1\dot{H}-\frac{r_2}{6},
\eea
whose solution is $H=\sqrt{\frac{r_1}{2}}\tanh(\sqrt{\frac{r_1}{2}}t)$ which just implies hyperbolic dependence of $a(t)$ (\ref{cosh}). It is well known that namely such a scenario (decreasing of scale factor changing then to increasing) is called bouncing scenario. We also introduce $h_1=\ddot{H}/M^3$.

The density can be found as well: if we use $G=M_P^{-2}$, and redefine $F(\Box)\to F(\Box/M^2)$, with $M$ is the characteristic nonlocality scale, we find
\bea
\rho_0=\frac{3(M^2_Pr_1-2\lambda f_0r_2)(r_2-12h_1M^4)}{12r^2_1-4r_2}.
\eea
Let us discuss possible implications of the equation (\ref{solh}). The cosmological constant turns out to be equal to $\Lambda=-\frac{r_2M^2_P}{4r_1}$, and there are three scenarios for evolution of the Universe:

1. $\Lambda<0$, $r_1>0$, $r_2>0$ -- cyclic Universe (in particular one can have cyclic inflation).

2. $\Lambda>0$, $r_1<0$, $r_2>0$ -- first contraction, then very rapid inflation (super-inflation) $a(t)\propto\exp(kt^2)$.

3. $\Lambda>0$, $r_1>0$, $r_2<0$ -- constant curvature $R=4\frac{\Lambda}{M^2_P}$, i.e. de Sitter solution.

So we find that accelerating solutions are possible within all these scenarios. Again, we note that in the constant scalar curvature case, we have drastic reducing of equations.

Moreover, it has been shown in \cite{Koshel} that for ${\cal L}=\sqrt{|g|}\sqrt{R-2\Lambda}F(\Box)\sqrt{R-2\Lambda}$, with $F(\Box)$ being an arbitrary analytic function, there are hyper-exponentially accelerating cosmological solutions $a(t)\propto e^{kt^2}$. 

The next step in study of nonlocal theories consists in introducing non-analytic functions of the d'Alembertian operator. The simplest case is $F(\Box)=\frac{1}{\Box}$. Actually it means that we must consider terms like $R\Box^{-1}R$. It is clear that the gravity extension with such a term is non-renormalizable since the propagator behaves as only $\frac{1}{k^2}$, so we gain nothing in comparison with the usual Einstein-Hlbert gravity \cite{Rach}. However, theories with negative degrees of the d'Alembertian operator can display new tree-level effects, especially within the cosmological context where an important class of nonlocal gravity models has been introduced in \cite{DWood}. The action of this class of theories is
\bea
\label{eliz}
S=\int d^4x\sqrt{|g|}\Big(\frac{1}{2G}\Big(R+Rf(\Box^{-1}R)-2\Lambda\Big)+{\cal L}_m\Big).
\eea
We note that the presence of the factor $\Box^{-1}$ actually implies in "retarded" solutions behaving similarly to the potential of a moving charge in electrodynamics. Further, this action has been considered in \cite{Eliz}, and below, we review the discussion given in that paper.

It is convenient to rewrite the action (\ref{eliz}) with use of two extra scalar fields $\xi$ and $\eta$:
\bea
\label{elscal}
S=\int d^4x\sqrt{|g|}\Big[\frac{1}{2G}[R(1+f(\eta)-\xi)+\xi\Box\eta-2\Lambda]+{\cal L}_m
\Big].
\eea
Varying this action with respect to $\xi$ and expressing $\eta=\Box^{-1}R$, we return to (\ref{eliz}). This corroborates the already mentioned idea that the modified gravity is in many cases equivalent to a some scalar-tensor gravity.

Then, we vary (\ref{elscal}) with respect to the metric and $\eta$ respectively:
\bea
&&\Box\xi+f_{\eta}(\eta)R=0;\nonumber\\
&&\frac{1}{2}g_{\mu\nu}[R(1+f(\eta)-\xi)-\pa_{\alpha}\xi\pa^{\alpha}\eta-2\Lambda]-R_{\mu\nu}(1+f(\eta)-\xi)+\nonumber\\
&+&\frac{1}{2}(\pa_{\mu}\xi\pa_{\nu}\eta+\pa_{\mu}\eta\pa_{\nu}\xi)-(g_{\mu\nu}\Box-\nabla_{\mu}\nabla_{\nu})(f(\eta)-\xi)=-GT_{\mu\nu}.\eea
We consider the FRW cosmological metric (\ref{FRW}) with $k=0$. 
As usual, the Hubble parameter is $H=\frac{\dot{a}}{a}$. The evolution equation for matter is usual: 
\bea
\dot{\rho}=-3H(\rho+p).
\eea
For scale factor and scalars, we have
\bea
&&2\dot{H}(1+f(\eta)-\xi)+\dot{\xi}\dot{\eta}+(\frac{d^2}{dt^2}-H\frac{d}{dt})(f(\eta)-\xi)+
G(\rho+p)=0;\nonumber\\
&&\ddot{\eta}+3H\dot{\eta}=-6(\dot{H}+2H^2);\nonumber\\
&&\ddot{\xi}+3H\dot{\xi}=-6(\dot{H}+2H^2)f_{\eta}(\eta).
\eea
We start with the de Sitter space corresponding to $H=H_0=const$, with the scalar curvature is $R=12H_0^2$. The equation of state is $p=\omega\rho$, as usual, so, we have the following solutions for the scalar $\eta$ and the density:
\bea
\label{sol}
\eta(t)&=&-4H_0(t-t_0)-\eta_0e^{-H_0(t-t_0)};\nonumber\\
\rho(t)&=&\rho_0e^{3(1+\omega)H_0t}.
\eea
Then we introduce the new variable $\Psi=f(\eta)-\xi$, and its equation of evolution is
\bea
\label{psi}
\ddot{\Psi}+5H_0\dot{\Psi}+6H^2_0(1+\Psi)-2\Lambda+G(\omega-1)\rho=0.
\eea
For $\eta$ we have
\bea
\label{eta}
\dot{\eta}^2f_{\eta\eta}+(\ddot{\eta}+3H_0\dot{\eta}-12H^2_0)f_{\eta}=\ddot{\Psi}+3H_0\dot{\Psi}.
\eea
This equation is a necessary condition for existence of the de Sitter solution.

Let us consider the particular case $\eta_0=0$ in (\ref{sol}). So, (\ref{eta}) reduces to
\bea
16H^2_0f_{\eta\eta}-24H^2_0f_{\eta}=\ddot{\Psi}+3H_0\dot{\Psi}.
\eea
So, knowing $\Psi$, one can find $f(\eta)$. It remains to solve (\ref{psi}). Some characteristic cases are:
\begin{itemize}
\item $\rho_0=0$: $\Psi=C_1e^{-3H_0t}+C_2e^{-2H_0t}-1+\frac{\Lambda}{3H^2_0}$;
\item $w=0$: $\Psi=C_1e^{-3H_0t}+C_2e^{-2H_0t}-1+\frac{\Lambda}{3H^2_0}-\frac{G\rho_0}{H_0}e^{-3H_0t}t$.
\item $w=-1/3$: $\Psi=C_1e^{-3H_0t}+C_2e^{-2H_0t}-1+\frac{\Lambda}{3H^2_0}+\frac{4G\rho_0}{3H_0}e^{-2H_0t}t$.
\end{itemize}
As for the function $f(\eta)$, in all cases it will be proportional to $e^{\eta/\beta}$, with $\beta>0$ (or, at most, linear combination of such functions with various values of $\beta$). Effectively we demonstrated arising of the exponential potential widely used in cosmology.

An important particular case is $\eta_0=0$. It follows from (\ref{sol}) that we have for $\beta\neq 4/3$:
\bea
\xi&=&-\frac{3f_0\beta}{3\beta-4}e^{-H_0(t-t_0)/\beta}+\frac{c_0}{3H_0}e^{-3H_0(t-t_0)}-\xi_0;\nonumber\\
\eta&=&-4H_0(t-t_0);\quad\, \omega=\frac{4}{3\beta}-1,\quad\, \Lambda=3H^2_0(1+\xi_0);\nonumber\\
\rho_0&=&\frac{6(\beta-2)H^2_0f_0}{\beta G},
\eea
so we can have exotic matter  for $0<\beta<2$. And at $\beta=2$ we have vacuum. If $\beta=4/3$, we have $\omega=0$, and $\rho<0$ (ghost-like dust).

However, we note that the nonlocal modifications of gravity are used mostly in cosmology. One of a few discussions of other metrics within the nonlocal gravity has been presented in \cite{Modesto3} where not only cosmological but also (anti) de Sitter-like solutions were discussed for theories involving, besides of.already mentioned term $R F(\Box)R$, also the terms $R_{\mu\nu} F_1(\Box)R^{\mu\nu}$ and $R_{\mu\nu\alpha\beta}F_2(\Box)R^{\mu\nu\alpha\beta}$, with $F,F_1,F_2$ are some functions of the covariant d'Alembertian operator.

Let us say a few words about other non-analytic nonlocal extensions of gravity. In \cite{Magg}, the additive term $\mu^2R\Box^{-2}R$ was introduced and shown to be consistent with cosmological observations. However, this theory turns out to be problematic from the causality viewpoint \cite{Zhang}. Also, in \cite{Ferna}, the first-order correction in $\mu^2$ to the Schwarzschild solution in a theory with this term has been obtained explicitly.

To close the discussion, it is important to note that the nonlocal gravity can arise as an effective theory as a result of integration over some matter fields. Namely in this manner, the term $R\Box^{-1}R$ contributes to the trace anomaly, at least in two dimensions, in \cite{AM}. Therefore, the presence of nonlocal terms can be apparently treated as a consequence of some hidden couplings with matter.

\section{Conclusions}

We discussed various nonlocal extensions of gravity. The key property of nonlocal theories is the possibility to achieve UV finiteness for an appropriate choice for nonlocal form factor(s). However, apparently explicit quantum calculations in nonlocal gravity models would be extremely complicated from the  technical viewpoint, therefore, up to now, all studies of such theories are completely classical ones. Moreover, most papers on nonlocal gravity models are devoted to cosmological aspects of these theories, and the results demonstrated along this chapter allow to conclude that nonlocal extensions of gravity can be treated as acceptable solutions for the dark energy problem. At the same time, nonlocal theories, including gravitational ones, display certain difficulties. The main problem is that one of unitarity and causality which still requires special attention. 

To conclude this chapter, let us emphasize the main directions for studies of nonlocal gravity models. First, clearly, it will be very important to check consistency of different known GR solutions, especially, various black holes (including f.e. non-singular and rotating ones). Second, various nonlocal form factors, not only Gaussian ones, are to be introduced, and their impact must be tested within the gravity context. Third, study of quantum effects in nonlocal gravity models is of special importance since namely at the perturbative level the main advantages of these theories such as the expected UV finiteness are crucial. It is natural to hope that these studies will be performed in next years.
 
\chapter{Summary} 
 
 We discussed various modifications of gravity introduced within the framework of the metric formalism. As we noted, in principle there are two fundamental problems to be solved by desired modifications of gravity: first, explanation of the cosmic acceleration, second, development of a theory consistent from the quantum viewpoint. Within the models we presented, different attempts to solve these problems are taken. It turns out to be that the problem of cosmic acceleration is solved by many extensions of gravity, and actually the main issue in this context consists in finding the theory fitting better the observational results (for discussion of cosmological constraining of gravitational models, see f.e. \cite{Melville} and many other papers). At the same time, the problem of formulating a perturbatively consistent gravity theory appears to be much more complicated. While the simplest way to construct the renormalizable gravity model is based on introducing higher-derivative terms, this manner suffers from the problem of arising ghost states. To solve the problem of ghosts, one can follow two ways: either break Lorentz symmetry in a strong manner introducing the HL gravity (effectively it means that we have a higher-derivative regularization in a spatial sector only) paying a price of arising a very complicated theory, and moreover, treating the Lorentz symmetry as an essentially low-energy phenomenon, or introduce nonlocality which allows to achieve renormalizability or even to rule out divergences, but, in this case, solving the problems of unitarity and causality would require special efforts. 
 
 One more way to solve the problem of renormalizability of the gravity is based on its supersymmetric extension. As it is well known, supersymmetric extension of any field theory improves essentially its ultraviolet behavior since so-called "miraculous cancellations" of UV divergences occur \cite{HST}. It is well known that the mechanism of these cancellation is very simple -- since fermionic contributions carry an extra minus sign, under an appropriate relations between coupling constants occurring due to the supersymmetry, some of fermionic divergent contributions cancel bosonic divergent contributions (for example, while the $\phi^4$ theory and Yukawa model display quadratic divergences, the Wess-Zumino model involving these theories as ingredients displays only logarithmic divergences). Moreover, there are known examples of completely finite supersymmetric theories, the paradigmatic example is the ${\cal N}=4$ super-Yang-Mills theory, where ${\cal N}$ is a number of supersymmetries (number of sets of generators of supersymmetry). Clearly, this called interest to a possible supersymmetric extension of gravity, so, the supergravity (SUGRA) was introduced (see \cite{Nieu} for a review). However, the ${\cal N}=1$ SUGRA is still non-renormalizable, therefore, the extensions of SUGRA with larger values of ${\cal N}$ began to be introduced. The maximal ${\cal N}$ allowing for a consistent theory is 8, for SUGRA (for larger values of ${\cal N}$, higher spin fields arise, and they cannot be consistently coupled to gravity. It should be noted also that the interest to SUGRA models with high ${\cal N}$ is motivated also by possible applications of these theories to superstrings. 
 
 So, let us briefly review the most important results found within ${\cal N}=8$ SUGRA obtained in series of papers by Bern, Dixon, Kosower and collaborators. In \cite{Bern1} it was proved that the degree of divergence, at ${\cal N}=8$, in $D$ dimensions and $L$ loops, is
\bea
\omega= (D-2)L-10.
\eea
 So we see that divergences in four dimensions can begin only from five-loop order! It is interesting to note that the approach from the same paper allows to show that the ${\cal N}=4$ super-Yang-Mills theory is all-loop finite.
 
 Further, on the base of the unitarity cuts approach, in \cite{Bern2}, it has been proved that the four-point functions in ${\cal N}=8$ SUGRA satisfies the same finiteness condition in the $D$-dimensional space-time
 \bea
 D<\frac{6}{L}+4,
 \eea
which for $D=4$ implies all-loop finiteness of these functions. Then, in \cite{Bern3}, with use of some identities applied for sets of more than 30 supergraphs, it was proved that some extra cancellations occur, so, the finiteness of ${\cal N}=8$ SUGRA is achieved up to four loops at $D\leq 5$. Afterwards, in \cite{Bern4} it was proved that the five-loop correction in this theory begins to diverge at $D\geq 24/5$, so, in the four-dimensional space-time, the theory is five-loop finite. Taking all together, we conclude that there is a natural hope that $N=8$ SUGRA is all-loop finite in $D=4$. The next problem consists in extracting some observable results for SUGRA (scattering amplitudes, corrections to GR etc.) while, up to now, there are only some isolated conclusions.

We conclude out course with the ideas that, first, in study of gravity one still has more questions that answers, second, apparently the most promising extensions of gravity are the SUGRA, the nonlocal gravity and the HL gravity. However, each of these modifications still has its difficulties which need to be solved. In principle, there are some other approaches to gravity, for example, treating the gravity as an emergent phenomenon caused by essentially quantum effects \cite{Verlinde}, asymptotic safety also known as non-perturbative renormalizability, which allows to treat many divergences as nonphysical ones \cite{Reuter}, bimetric gravity based on use of the additional second-rank symmetric tensor, Palatini approach treating metric and connection as independent variables, and, clearly, modifications of gravity based on use of non-Riemannian geometry, especially, torsion and nonmetricity. To finish, we note again that in gravity there is still much more questions than answers.
 
\newpage

{\bf Acknowledgements.} Author is grateful to Profs.  M. Gomes,  J. R. Nascimento,   A. J. da Silva,  T. Mariz, G. Olmo, A. F. Santos, E. Passos, P. Porfirio  for fruitful collaboration and interesting discussions. The work has been partially supported by CNPq, project 301562/2019-9.

\end{document}